%% file: main.tex
\PassOptionsToPackage{x11names,table,xcdraw}{xcolor}
\documentclass[acmsmall,natbib=false,authorversion]{acmart}
\usepackage{color}

\AtBeginDocument{%
    }

\setcopyright{acmlicensed}
\acmJournal{TIIS}
\acmYear{2024} \acmVolume{1} \acmNumber{1} \acmArticle{1} \acmMonth{1}
\acmDOI{10.1145/3652028}

\acmJournal{JACM}
\acmVolume{37}
\acmNumber{4}
\acmArticle{111}
\acmMonth{8}

\RequirePackage[
    datamodel=acmdatamodel,
    style=acmauthoryear
]{biblatex}

\input{config}

\title[\inlinegraphicsm{1em}{-0.12em}{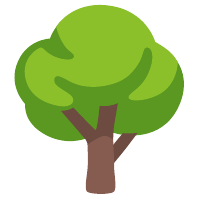}-\textbf{generAItor:} Tree-in-the-Loop Text Generation for Language Model Explainability and Adaptation]{\inlinegraphicsm{1em}{-0.12em}{images/googlefonts_noto-emoji/emoji_u1f333}-\textbf{generAItor:}  Tree-in-the-Loop Text Generation\\ for Language Model Explainability and Adaptation}
\cleantitle{generAItor: Tree-in-the-Loop Text Generation for Language Model Explainability and Adaptation}

\author{Thilo Spinner}
\affiliation{%
    \institution{ETH Zurich}
    \city{Zurich}
    \country{Switzerland}}
\email{thilo.spinner@inf.ethz.ch}

\author{Rebecca Kehlbeck}
\affiliation{%
    \institution{University of Konstanz}
    \city{Konstanz}
    \country{Germany}}
\email{rebecca.kehlbeck@uni-konstanz.de}

\author{Rita Sevastjanova}
\affiliation{%
    \institution{ETH Zurich}
    \city{Zurich}
    \country{Switzerland}}
\email{rita.sevastjanova@inf.ethz.ch}

\author{Tobias Stähle}
\affiliation{%
    \institution{University of Konstanz}
    \city{Konstanz}
    \country{Germany}}
\email{tobias.staehle@uni-konstanz.de}

\author{Daniel A.\ Keim}
\affiliation{%
    \institution{University of Konstanz}
    \city{Konstanz}
    \country{Germany}}
\email{keim@uni-konstanz.de}

\author{Oliver Deussen}
\affiliation{%
    \institution{University of Konstanz}
    \city{Konstanz}
    \country{Germany}}
\email{oliver.deussen@uni-konstanz.de}

\author{Mennatallah El-Assady}
\affiliation{%
    \institution{ETH Zurich}
    \city{Zurich}
    \country{Switzerland}}
\email{menna.elassady@ai.ethz.ch}

\renewcommand{\shortauthors}{Spinner et al.}

\begin{document}

\begin{abstract}
    \input{sections/abstract}
\end{abstract}



\received{18 July 2023}
\received[revised]{16 November 2023 and 26 January 2024}
\received[accepted]{30 January 2024}

\maketitle

\clearpage
\section{Introduction}
\label{sec:introduction}
\input{sections/introduction}

\section{Related Work}
\label{sec:related-work}
\input{sections/related-work}

\section{Problem Characterization}
\label{sec:problem-characterization-and-methodology}
\input{sections/problem-characterization}

\section{Tree Visualization \& Model Configuration}
\label{sec:beam-search-tree}
\input{sections/tree-visualization-and-model-configuration}

\section{Text Generation, Comparison, \& Model Adaptation}
\label{sec:task-specific-widgets}
\input{sections/text-generation-comparison-and-model-adaptation}

\section{Evaluation}
\label{sec:evaluation}
\input{sections/evaluation}

\section{Discussion}
\label{sec:discussion}
\input{sections/discussion}

\section{Conclusion}
\label{sec:conclusion}
\input{sections/conclusion}

\clearpage
\printbibliography

\clearpage
\appendix

\section{Natural Language Processing Pipelines}
\label{apx:sec:pipelines}
\input{appendix/pipelines}

\end{document}

%% file: config.tex
\usepackage{xurl} 
\usepackage{multirow}
\usepackage{colortbl}
\usepackage{tabularx}
\usepackage{tabulary}
\usepackage{enumitem}
\usepackage{amsfonts}
\usepackage{amsmath}
\usepackage[nameinlink,capitalise]{cleveref}
\usepackage{calc}
\usepackage{changepage}
\usepackage{wrapfig}
\usepackage{subcaption}
\usepackage{ragged2e}
\usepackage{lipsum}
\usepackage[normalem]{ulem}
\usepackage{overpic}
\usepackage[many]{tcolorbox}
\usepackage{listings}
\usepackage{siunitx}
\usepackage{adjustbox}
\usepackage{xspace}
\usepackage{pdflscape}
\usepackage{tikz}
\usepackage{svg}
\usepackage{mdframed}
\usepackage{array}
\usepackage{varwidth}
\usepackage{soul}
\usepackage{titlesec}

\newcommand{\bias}[1]{{\leavevmode\color{Red4}#1}}

\crefname{table}{table}{tables}
\crefname{figure}{figure}{figures}
\crefname{equation}{equation}{equations}
\crefname{part}{part}{parts}
\crefname{section}{section}{sections}
\crefname{subsection}{section}{section}
\crefname{subsubsection}{section}{section}
\Crefname{lstlisting}{algorithm}{algorithms}

\Crefname{table}{Table}{Tables}
\Crefname{figure}{Figure}{Figures}
\Crefname{equation}{Equation}{Equations}
\Crefname{part}{Part}{Parts}
\Crefname{section}{Section}{Sections}
\Crefname{subsection}{Section}{Sections}
\Crefname{subsubsection}{Section}{Sections}
\Crefname{lstlisting}{Algorithm}{Algorithms}

\titleformat{\subsubsection}
{\normalfont\normalsize\itshape}{\thesubsubsection}{1em}{}
\titlespacing*{\subsubsection}{0pt}{3.25ex plus 1ex minus .2ex}{0ex plus .2ex}

\makeatletter
\newcommand{\cleantitle}[1]{\def\@cleantitle{#1}}
\AtBeginDocument{\hypersetup{pdftitle=\@cleantitle}}
\makeatother
\hypersetup{%
    pdftitle={TEMP}%
}

\setdescription{labelindent=1.5em,leftmargin=3em}

\definecolor{MiroBlack}{HTML}{1a1a1a}
\definecolor{PredictionPipelineColor}{HTML}{7AB635}
\definecolor{KwEmbeddingPipelineColor}{HTML}{3C9CDB}
\definecolor{BabelnetEmbeddingPipelineColor}{HTML}{F24726}
\definecolor{OntoReplacePipelineColor}{HTML}{0CA789}
\colorlet{SummaryBoxColor}{Snow3}

\makeatletter
\newtcbox{\kwColorBox}[1][]{on line,fontupper=\footnotesize\sffamily\bfseries\small,boxrule=0.5pt,arc=2pt,coltext=#1,colback=#1!10!white,colframe=#1,boxsep=0pt,left=1.5pt,right=1.5pt,top=1.5pt,bottom=1.5pt}
\makeatother

\newcommand{\kw}[2]{%
    \begin{kwColorBox}[#2]%
    {#1}%
    \end{kwColorBox}%
    \xspace%
}

\renewcommand{\paragraph}[1]{\refstepcounter{paragraph}\noindent\textbf{#1\ ---}\label{par:\theparagraph}}

\definecolor{ColorUser}{HTML}{A42C2C}
\definecolor{ColorChallenge}{HTML}{A02FA5}
\definecolor{ColorTask}{HTML}{408E2F}
\definecolor{ColorWidget}{HTML}{4F85C3}
\newcommand{\rawuserdonotuse}[1]{\kw{#1}{ColorUser}}
\newcommand{\defuser}[1]{\rawuserdonotuse{\phantomsection\label{user:#1}#1}}
\newcommand{\refuser}[1]{\hyperref[user:#1]{\rawuserdonotuse{#1}}}

\newcommand{\rawchallengedonotuse}[1]{\kw{#1}{ColorChallenge}}
\newcommand{\defchallenge}[1]{\rawchallengedonotuse{\phantomsection\label{challenge:#1}#1}}
\newcommand{\refchallenge}[1]{\hyperref[challenge:#1]{\rawchallengedonotuse{#1}}}

\newcommand{\rawtaskdonotuse}[1]{\kw{T#1}{ColorTask}}
\newcommand{\deftask}[1]{\rawtaskdonotuse{\phantomsection\label{task:#1}#1}}
\newcommand{\reftask}[1]{\hyperref[task:#1]{\rawtaskdonotuse{#1}}}

\newcommand{\blankwidget}{\kw{W}{ColorWidget}}
\newcommand{\rawwidgetdonotuse}[1]{\kw{W#1}{ColorWidget}}
\newcommand{\defwidget}[2]{\rawwidgetdonotuse{\phantomsection\label{widget:#1}#2}}
\newcommand{\refwidget}[2]{\hyperref[widget:#1]{\rawwidgetdonotuse{#2}}}

\mdfdefinestyle{leftline}{
    linewidth=1pt,
    linecolor=gray!50,
    leftline=true,
    rightline=false,
    topline=false,
    bottomline=false,
    backgroundcolor=white,
    skipabove=3pt, 
}
\newcommand{\taskbox}[2]{
    \begin{minipage}{\linewidth}
        \vspace{0.25em}
        \begin{tabularx}{\linewidth}{ >{\columncolor{ColorTask!30!white}}m{0.5cm} >{\columncolor{ColorTask!10!white}}X }
            #1 & #2
        \end{tabularx}
        \vspace{0.25em}
    \end{minipage}
}

\definecolor{ColorBSPrimary}{HTML}{2C3E50}

\newcommand{\HEAD}{\textcolor{ColorBSPrimary}{\textsc{Head}}\xspace}

\newlength\myheight%
\newlength\mydepth%
\settototalheight\myheight{Xygp}%
\newcommand*\inlinegraphics[1]{%
    \settototalheight\myheight{Xygp}%
    \settodepth\mydepth{Xygp}%
    \raisebox{-\mydepth}{\includegraphics[height=\myheight]{#1}}%
}%

\newcommand*\inlinegraphicsm[3]{%
    \raisebox{#2}{\includegraphics[height=#1]{#3}}%
}%

%% file: sections/abstract.tex
Large language models (LLMs) are widely deployed in various downstream tasks, e.g., auto-completion, aided writing, or chat-based text generation.
However, the considered output candidates of the underlying search algorithm are under-explored and under-explained.
We tackle this shortcoming by proposing a \textit{tree-in-the-loop} approach, where a visual representation of the beam search tree is the central component for analyzing, explaining, and adapting the generated outputs.
To support these tasks, we present generAItor, a visual analytics technique, augmenting the central beam search tree with various task-specific widgets, providing targeted visualizations and interaction possibilities.
Our approach allows interactions on multiple levels and offers an iterative pipeline that encompasses generating, exploring, and comparing output candidates, as well as fine-tuning the model based on adapted data.
Our case study shows that our tool generates new insights in gender bias analysis beyond state-of-the-art template-based methods.
Additionally, we demonstrate the applicability of our approach in a qualitative user study.
Finally, we quantitatively evaluate the adaptability of the model to few samples, as occurring in text-generation use cases.

%% file: sections/introduction.tex
Recently, large language models (LLMs) have gained increased popularity, especially in the field of natural language generation (NLG).
At the latest, with the introduction of ChatGPT\footnote{\url{https://openai.com/blog/chatgpt}}, LLMs have been made accessible to a wider, more general audience.
However, despite their growing recognition and notable accomplishments, they still face several limitations.
Common failures, even for state-of-the-art models, are repetitive content, the lack of factual accuracy, often referred to as hallucination~\cite{Ji2023SurveyHallucinationNatural}, and biases~\cite{Alba2022OpenaiChatbotSpits}.
However, the perceived high quality of LLM outputs makes identifying errors in their predictions difficult, which is aggravated by a lack of explainability and accessibility~\cite{Zhao2024ExplainabilityLargeLanguage}.
Gaining understanding and access to the model’s decision-making process is fundamental for recognizing errors in their outputs, calming concerns about overestimating the model’s capabilities, and empowering users to guide the model’s predictions to align with their intentions.
Particularly, the chat interface of ChatGPT and other chat- or completion-based approaches omit important information on uncertainties or viable alternatives from the users.
While text-based interfaces may fulfill the needs for a broad, general audience, \textbf{interested non-experts} and \textbf{linguistic experts} require more in-depth insights and control.

We identify three primary shortcomings in the current state-of-the-art for interacting with LLMs: lack of \textbf{explainability}, \textbf{comparability}, and \textbf{adaptability}.
Explainability refers to understanding of the model's decision process, including the way a language model predicts its output, its sampling strategy, and the probabilities of these outputs.
For example, explanations of a prediction's certainty can provide the user a hint on possible hallucinations.
Comparability, i.e., a simple yet effective comparison of multiple generated outputs, can enable the user to assess more specific nuances in the model's predictions.
This kind of contrastive explanation~\cite{ElAssady2019TowardsXaiStructuring} is particularly relevant for linguistic experts.
For instance, by adapting prompts with typical names from varying ethnic groups and comparing the predictions, the user can assess the model's biases, if present.
And lastly, adaptability is relevant when the generated output is not satisfactory.
The insights gained from explainability and comparability empower the user to steer the model towards their intentions.
Concretely, the user should be able to edit problematic parts; e.g., by correcting made-up facts and making these changes permanent; e.g., by fine-tuning the model.

Since almost all modern LLMs have committed themselves to the transformer architecture, besides their number of trainable parameters, the quality of the training data is the decisive factor for a model's performance~\cite{Lauscher2021SustainableModularDebiasing,Mishra2022CrossTaskGeneralization}.
Therefore, studying the model's behavior is closely linked to studying its in-- and outputs, representing a local approximation of the information the model has learned during training.
Our proposed approach, thus, focuses on making these in-- and outputs accessible and explorable to the user.
A straightforward way to achieve this is to make the search algorithm transparent.
The most prominent algorithm to sample sequences from the probability distributions output by the model is \textit{beam search}.
By sampling the \textit{decision-space}~\cite{ElAssady2018ThreadreconstructorModelingReply} through expanding the most promising sequence in a limited set of candidate sequences, the algorithm results in a tree, scanning the search space for sequences with high overall probability.
Beam search is thus commonly used in language model explanation methods, such as the visual interface by Lee et al.~\cite{Lee2017InteractiveVisualizationManipulation}, Seq2Seq-Vis~\cite{Strobelt2018Seq2seqVisVisual}, or GenNI~\cite{Strobelt2022GenniHumanAi}.

In this paper, we propose a \textit{tree-in-the-loop} interaction paradigm, which leverages a visual representation of the \textit{beam search tree} (BST) as the central component of the \textbf{generAItor} visual analytics technique.
We reveal and explain the model's decision-making process by laying out the BST and augmenting it with additional explanations, such as token probabilities, semantic keyword coloring, and sentiment annotations.
Comparative explanations are facilitated by juxtaposing multiple BSTs, allowing the user to compare the model's predictions under slightly varied inputs.
Furthermore, we enable the user to interact with the tree, allowing them to adapt and steer the model's predictions, for example, by overriding model decisions, editing predicted sequences, or fine-tuning the model.
To facilitate an effective analysis through visual interactive methods,
we identify five main tasks in the context of informed text generation: model prompting and configuration, tree exploration and explainability, guided text generation, comparative analysis, and BST and model adaptation.
Each of these tasks places distinct demands on the tools available.

To be able to fulfill these demands in a combined approach, we design \textbf{a modular, widget-based workflow}, where task-specific widgets enhance the BST with tailored controls, interaction possibilities, and visualizations.
Each widget adds a very specific functionality.
However, in symbiosis, a selected set of task-supporting widgets, in interaction with the search tree, enables novel, powerful modes of analysis.
E.g., comparative analysis is facilitated by two particular widgets, allowing linguistic experts to observe changes in the tree under slight variations of the starting prompt.
This reveals biases in the observed model, whose identification and mitigation is one of the most burning issues with state-of-the-art language models~\cite{Alba2022OpenaiChatbotSpits}.

\noindent\textbf{In this paper, we contribute:}
(1) A detailed problem analysis of the challenges of explainability, controllability, and adaptability in the context of various text generation tasks.
(2) A novel visual analytics technique called generAItor, tackling these challenges in an interactive tree-in-the-loop-approach.
(3) An implementation of the generAItor technique in a web-based visual analytics workspace.
(4) A three-fold evaluation of the generAItor technique, including (4.1) case studies, showcasing the generative and comparative capabilities of our technique, (4.2) a qualitative user-study, proving the usability of the implementation, and (4.3) a quantitative evaluation, confirming the ability to adapt the model to user-preferences with few training samples.

%% file: sections/related-work.tex
In the following, we present our related work on language modeling, semantic similarity, controlled text generation, and bias analysis.

\subsection{Language Modeling}
Language models (LMs) are probability distributions over word sequences and a core component of natural language processing (NLP) systems~\cite{Bengio2000NeuralProbabilisticLanguage}.
With the emergence of the transformer architecture~\cite{Vaswani2017AttentionIsAll}, there was a paradigm shift away from recurrent neural networks~\cite{Rumelhart1986LearningRepresentationsBack} since transformers allow parallel computations, speeding up training times, and prove superior in capturing long-term dependencies~\cite{Vaswani2017AttentionIsAll}.
They use the attention mechanism~\cite{Bahdanau2014NeuralMachineTranslation}, which directs the focus on important tokens in the input sequence.
Nowadays, numerous pre-trained transformer architectures are available for public use~\cite{Wolf2020TransformersStateArt}.
There are different types of transformers, whereby the two main categories are masked language models and generative language models.

\paragraph{Masked LMs}
BERT~\cite{Devlin2018BertPreTraining} is a transformer-based LM that was trained on masked language modeling (i.e., \textit{cloze}) and next-sentence prediction tasks and is commonly fine-tuned for diverse text classification tasks~\cite{Howard2018UniversalLanguageModel}.
Due to its pre-training objective, BERT (as well as other masked language models) is not suitable for text generation tasks.
We use BERT for masked word prediction in the \emph{ontological replace} functionality~\refwidget{OR}{\inlinegraphicsm{1em}{-0.1em}{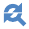}}.

\paragraph{Generative LMs}
Text can be generated using generative transformer models, such as GPT-2~\cite{Radford2019LanguageModelsAre}, GPT-3~\cite{Brown2020LanguageModelsAre}, or GPT-4~\cite{OpenAI2023Gpt4Technical}.
These are autoregressive models that were pre-trained on the causal language modeling task, learning to predict the next word in the input sequence.
For a broader overview, see the survey on pre-trained language models for text generation by Lie et al.~\cite{Li2021PretrainedLanguageModel}.
In our work, we use GPT-2 and Bloom~\cite{Workshop2023Bloom176bParameter} for text generation; however, the approach is designed to support other transformer-based LMs as well.

\subsection{Semantic Similarity}

\paragraph{Word Taxonomies and Ontologies}
Leveraging semantic graphs and knowledge bases, such as YAGO and DBpedia, it is possible to infer concept or topic hierarchies via language models~\cite{Chen2021ConstructingTaxonomiesPretrained,Huang2020CorelSeedGuided,Zhang2018Taxogen} or expand existing taxonomies ~\cite{Jiang2022TaxoenrichSelfSupervised,Xu2022TaxopromptPromptBased}.
Methods such as OntoEA~\cite{Xiang2021OntoeaOntologyGuided} align entities by jointly embedding ontologies and knowledge bases.
Taxonomies can be used to improve recommender systems~\cite{Tan2022EnhancingRecommendationAutomated} and help with entity recognition~\cite{Li2022Taxotrans} or translation~~\cite{Li2022Taxotrans}.
WordNet information can be integrated into pre-trained language models for improved sense disambiguation, e.g., ARES~\cite{Scarlini2020MoreContextsComes}, or used to build human-readable concept vectors~\cite{Conia2020ConceptionMultilinguallyEnhanced}.
For our method, we use ARES and BERT embeddings in conjunction to create domain-specific predictions with an ontology graph~\refwidget{OVM}{\inlinegraphicsm{1em}{-0.1em}{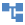}} created from the BabelNet~\cite{Navigli2012BabelnetAutomaticConstruction} semantic graph.

\paragraph{Embedding Similarity}
In language models, each token of the input text is mapped to a high-dimensional vector.
Related work has shown that these context-dependent embeddings encode different context/language properties.
Although BERT is the most widely analyzed language model so far~\cite{Rogers2020PrimerBertologyWhat}, other transformer models, such as GPT-2, and their produced embedding spaces have also attracted computational linguistics' and visual analytics researchers' attention~\cite{Ethayarajh2019HowContextualAre,Sevastjanova2022LmfingerprintsVisualExplanations}.
Prior research has shown that semantic information, such as word senses and semantic roles, is captured best in the higher layers of transformer models~\cite{Reif2019VisualizingMeasuringGeometry,Wiedemann2019DoesBertMake,Sevastjanova2022LmfingerprintsVisualExplanations}.
Thus, these contextualized embeddings are commonly used as features for semantic similarity tasks.
In our work, we apply a dimensionality reduction technique on embeddings extracted from the used LMs to map the tokens to unique colors based on their coordinates in the two-dimensional space.
With this approach, tokens with a semantic similarity get assigned to similar colors~\cite{ElAssady2022SemanticColorMapping}.

\subsection{Controlled Text Generation}

\paragraph{Algorithmic Approaches}
In general, controlling the style and information of natural language generation is one of the applications identified by Gatt and Krahmer~\cite{Gatt2018SurveyStateArt}.
One challenge of integrating knowledge into text generation is the automatic steering of the generation in a particular direction.
Using plug-and-play language models is one possibility to steer text generation~\cite{Qin2020BackFutureUnsupervised}.
Concerning pre-trained language models, it is possible to control, e.g., the sentiment~\cite{Dathathri2019PlugPlayLanguage,Hu2017ControlledGenerationText}, keywords~\cite{He2021ParallelRefinementsLexically}, or the topic~\cite{Dathathri2019PlugPlayLanguage}.
Frameworks such as FAIR~\cite{Hua2020PairPlanningIterative} allow the generation of content-controlled text by combining BERT with BART~\cite{Lewis2020BartDenoisingSequence}.
A larger overview is given in the survey by Zhang et al.~\cite{Zhang2022SurveyControllableText}.
Building on this, many approaches now integrate external resources such as knowledge bases.
More details can be found in the survey by Yu et al.~\cite{Yu2022SurveyKnowledgeEnhanced}.
However, these techniques do not allow immediate intervention in the decision process, which we specifically target with our approach.

\paragraph{Visual Interactive Approaches}
Focusing on interactive editing, \citeauthor{Du2022ReadReviseRepeat}~\cite{Du2022ReadReviseRepeat} provide interactive suggestions in their tool to achieve high-quality text edits with minimal human effort.
\citeauthor{Padmakumar2022MachineLoopRewriting}~\cite{Padmakumar2022MachineLoopRewriting} use a human-in-the-loop approach to replace text segments for the task of creative image captioning.
\citeauthor{Gehrmann2019VisualInteractionDeep}~\cite{Gehrmann2019VisualInteractionDeep} propose an interactive framework that allows users to control generative segments through a process called collaborative semantic inference.
Following this, \citeauthor{Strobelt2022GenniHumanAi}~\cite{Strobelt2022GenniHumanAi} create GenNi, an interface for collaborative text generation.
They guide the model output using explicitly defined constraints.
The user has to know beforehand how he wants to control the model output, as it is not possible to adapt the state during inference.
With Wordcraft, \citeauthor{Yuan2022WordcraftStoryWriting}~\cite{Yuan2022WordcraftStoryWriting} present an interactive interface that allows writers to create stories with the assistance of large language models.
Their system lets authors re-write, replace, and auto-generate text, as well as define custom requests to the language model.
In contrast, our approach enables direct interaction with the model's outputs by exposing predictions and probabilities in the beam search tree.

\subsection{Bias Analysis}
Current research explores not only what the models learn but also when they fail and which limitations they have, such as different types of biases~\cite{GarridoMunoz2021SurveyBiasDeep}.
For instance, Blodgett et al.~\cite{Blodgett2020LanguageTechnologyIs} present a taxonomy for fairness definitions that machine learning
researchers have defined to avoid existing bias in AI systems.
Mehrabi et al.~\cite{Mehrabi2021SurveyBiasFairness} define the bias problem specifically in language modeling tasks in a formal way and explore how it has been treated in related work
regarding their detection and correction.

In masked language models, the detection of bias is typically done by applying templates or pre-defined word lists.
For instance, the Word Embedding Association Test (WEAT)~\cite{Caliskan2017SemanticsDerivedAutomatically} measures the association between two target word sets (e.g., male pronouns and, e.g., female pronouns) based on their cosine similarity to words from two attribute sets (e.g., terms related to science or art) to make conclusions about encoded biases.
Liang et al.~\cite{Liang2021TowardsUnderstandingMitigating} show that the analysis of biases in text generation can be more nuanced, e.g., biases can arise during the generation of any token~\cite{Nadeem2021StereosetMeasuringStereotypical}.
Alnegheimish et al.~\cite{Alnegheimish2022UsingNaturalSentence} find that bias ``evaluations are very sensitive to the design choices of template prompts.''
According to the authors, the use of template-based prompts tends to evoke biases from the model's default behavior rather than reflecting the actual correlation between gender and profession, analyzed in their work.
Thus, we propose a tree-based approach for comparative, exploratory bias analysis, allowing the detection of biases in variable-length sequences and the identification of subtle nuances in the model's predictions.
For a detailed case study, show-casing the benefits of our comparative approach, see~\cref{subsec:case-study-comparative-analysis}.

%% file: sections/problem-characterization.tex
With recent advances in language generation and the release of ChatGPT, language models have made their way into mainstream use.
While automatic text generation through language models can support the author through corrections, suggestions, or chat-based question answering, understanding of the model's capabilities and limitations and access to its predictions is still limited.
However, such understanding and access are crucial for raising awareness of dangers (e.g., biased outputs, hallucinations), allaying fears of its potential (e.g., overestimation of a model's capabilities), and enabling users to steer the model's predictions towards their intention (e.g., by selecting or modifying outputs).

While the average user might not be willing to invest time and effort in investigating the behavior of language models, we identify two primary user groups with different interests and requirements for language model analysis.
We define \emph{non-experts} \defuser{Non} as interest-driven persons with an affinity for technical advancements and the wish to explore modern language models.
The term ``non-expert'' only refers to the user's experiences with large language models and their background in computational linguistics; they can still be domain experts in other fields.
Examples could be a journalist who writes about language models and wants to understand their capabilities and limitations or a writer who wants to use LLMs to generate text with a specific style or topic.
Complementary, we define \emph{linguistic experts} \defuser{Lin} as users working in (computational) linguistics, with a main focus on the analysis of model behavior.
An example could be a linguist who wants to observe biases encoded in the model~\cite{Spinner2023RevealingUnwrittenVisual}.
Our approach is targeted towards both user groups, with shifting focus on the tasks our system supports.
For the non-experts, understanding of the model's capabilities, exploration of outputs, investigation of uncertainties, and the ability to adapt model outputs are primarily important.
In contrast, the linguistic expert is interested in the close analysis of model outputs, e.g., to observe learned syntactic and semantic structures, identify model defects, or assess model biases.
In the following, we specify the challenges and tasks for the derived user groups.

\vspace{0.1cm} 

\subsection{Challenges}
\label{subsec:challenges}
The challenges are derived from research gaps in related work and from discussions with non-experts~\refuser{Non}, machine learning experts, and computational linguists~\refuser{Lin}.

\paragraph{Explainability \defchallenge{Ex}}
Despite the impressive performance of state-of-the-art language models, their predictions are often underexplained, as deep-learning-based models are typically black boxes, making explainability a major challenge~\cite{Danilevsky2020SurveyStateExplainable}.
However, language models have the advantage of interpretable in- and outputs (namely: text) and easy-to-understand prediction mechanisms, which we aim to leverage to solve this challenge.
We identify two primary aspects of explainability regarding language models: model and output explainability.
Explainability is important for both the non-expert \refuser{Non} and the linguistic expert \refuser{Lin}.

\emph{Model explainability} relates to explanations of the model's algorithmic approach, such as providing information on the model's architecture, the used search algorithm, or the influence of randomness (c.f., reproducibility)~\cite{Spinner2020ExplainerVisualAnalytics}.
Particularly, mainstream media often fail to explain the primary mechanism behind LLMs: predicting the likelihood of tokens to follow a sequence of previous tokens.
Although some articles briefly touch the topic~\cite{Metz2022NewChatbotsCould,Roose2023HowChatbotsLarge}, there is much misinformation through excessive abstraction and a lack of easy-to-follow visualizations and interactive systems that could impart a thorough understanding to non-experts.
Understanding this mechanism is crucial to raising awareness of a model's limitations and allaying fears of its potential.
\emph{Output explainability} refers to explanations of the model's token representations and output probabilities, such as token embedding similarity or output certainty.

\paragraph{Comparability \defchallenge{Com}}
The ability to explore the space of possible model outputs is vast and currently underexplored~\cite{Alnegheimish2022UsingNaturalSentence}.
For the analysis, instance-based comparability of generated outputs is essential for linguistics, e.g., for bias analysis or hypothesis generation.
Particularly, non-template based, explorative analysis enables hypotheses generation and inductive learning~\cite{Sternberg2016CognitivePsychology}.

\paragraph{Adaptability \defchallenge{Ada}}
Even state-of-the-art language models often fail to produce output which aligns with human intentions and sticks to facts~\cite{Ji2023SurveyHallucinationNatural,LeCun2023DoLanguageModels}.
Therefore, adaptability is essential to employ language models in real-world scenarios.
Again, we differentiate two sub-aspects: output adaptability and model adaptability.
\emph{Output adaptation} refers to direct edits of the model's predictions, e.g., to correct hallucinated facts, re-prime the model through entering custom text, or select from alternative outputs, targeting both the non-expert~\refuser{Non} and linguistic expert~\refuser{Lin}.
That followed, \emph{model adaptation} relates to model fine-tuning with the edited data to make changes permanent for future sessions, which is also relevant for both user groups.

\vspace{0.1cm} 

\subsection{The Tree-in-the-Loop Approach}

To address the challenges identified above, we propose the \textit{tree-in-the-loop} paradigm, a novel approach to interactively explore and adapt the predictions of language models through the visualization of the beam search tree.

With the invention of transformers, the architecture of state-of-the-art models is well-established, shifting the focus for performance improvements on the training process and the quality of training data~\cite{Ouyang2022TrainingLanguageModels}.
Consequently, understanding a model's behavior involves examining its inputs and outputs, which reflect the ``knowledge'' it has acquired during training.
Therefore, our approach emphasizes making these inputs and outputs more user-accessible and explorable.

In each step, when predicting the next token for a given input sequence, the model outputs a probability distribution over all known tokens.
The final text has to be constructed by sampling from this probability distribution.
A common heuristic to choose the output with the highest probability is beam search.
Beam search is a greedy search algorithm that expands the $k$ most likely sequences in each step, resulting in a tree with $k$ nodes in each tree level.
$k$ is called the \textit{beam width}.
Branches with low overall probability stall in this process, resulting in a tree with varying depth.
The deepest leaf node with the highest probability is then chosen as the final output.
Often, additional parameters are used to increase the diversity of the generated text, e.g., by penalizing the repetition of $n$-grams or by adding randomness to the sampling process, e.g., through top-$k$ sampling or temperature scaling.

Most interfaces only present the user with the final text, discarding all information about the sampling process, such as uncertainties of predictions, alternative outputs, or the influence of parameters such as the beam width or an $n$-gram penalty.
To enable an understanding of the model's prediction process, we aim to make this information accessible to the user.
This is most straightforwardly done by visualizing the beam search tree, which is easy to understand and interact with.
Furthermore, it provides a direct representation of the underlying sampling algorithm and thus does neither neglect information nor introduce false rationalization.

The tree-in-the-loop approach is the extension of the beam search tree with additional augmentations, visualizations, and interaction possibilities.
This makes the tree accessible to non-technical users~\refuser{Non} and supports linguistic experts~\refuser{Lin} in the advanced analysis of linguistic phenomena.

\subsection{User Tasks}
\label{subsec:tasks}

From the before-discussed challenges of explainability, adaptability, and comparability, we derive the following user tasks, as depicted in~\cref{fig:teaser}.
While some tasks are essential to load and interact with LLMs, others are optional and only relevant for specific use cases.

\vspace{0.25em}

\paragraph{Model Prompting and Configuration}
To choose and asses models from the vast zoo of pre-trained LLMs~\cite{Wolf2020TransformersStateArt}, the user has to be able to load different models.
Furthermore, the user should be able to provide a prompt to the model and configure parameters for the prediction algorithm.
After interactively editing outputs and, potentially, fine-tuning the model, the user should be able to save the refined sequences and model for future sessions.
Since these tasks describe basic interactions with the model, they are equally important for the linguistic expert \refuser{Lin} and the non-technical user \refuser{Non}.
\\
\taskbox{\deftask{0}}{
    Load and assess (pre-trained) models, provide prompts, and configure parameters for the prediction algorithm.
    Save trees and models for future sessions.
}

\paragraph{Tree Exploration \& Explainability}
The beam search tree, used to sample model outputs, should be transparent and accessible to the user, allowing them to explore alternatives and assess the certainty of the model's predictions, addressing the explainability challenge \refchallenge{Ex}.
Supporting beam search exploration, semantic annotations of the tree should be provided, e.g., to identify topic similarity or to discover undesired patterns like looping structures.
This is important for both the non-expert \refuser{Non} and for the linguistic expert \refuser{Lin}, who are interested in the close analysis of model outputs and need a higher-level overview to cover large trees.
\\
\taskbox{\deftask{1}}{
    Assess probabilities and explore alternative branches in the beam search tree.
    Identify topic similarity and undesired patterns, such as looping structures.
}

\paragraph{Guided Text Generation}
Using the start prompt or existing sequences from the tree, the user should be able to query the LLM to extend the beam search tree with new predictions.
Since the beam search tree might grow to a significant size, a text view should be provided to close-read generated text and navigate the beam search tree to a local context.
Also, for longer texts, an overview of the topics touched facilitates an overview and understanding of the generated text.
This task mainly targets the non-expert \refuser{Non}, who is likely to generate longer texts.
\\
\taskbox{\deftask{2}}{
    Query the LLM to extend the beam search tree.
    Navigate the beam search tree to a local context.
    Investigate the topics touched by the generated text and stalled beam search branches.
}

\paragraph{Comparative Analysis}
Comparative analysis tackles the comparability challenge \refchallenge{Com} and is particularly important for the linguistic expert \refuser{Lin}, who is interested in the close analysis of model outputs.
Different trees can be generated and compared by varying start prompt and beam search parameters, allowing to assess the effects of those changes.
Semantic annotations and aggregated representations should be provided to quickly identify the key differences between trees.
This facilitates, e.g., generating new hypotheses, analyzing model biases, or investigating the influence of function words on the predictions.
\\
\taskbox{\deftask{3}}{
    Generate and compare different trees by varying prompt and beam search parameters.
    Observe syntactic and semantic differences in the trees.
}

\paragraph{BST Adjustment \& Model Adaptation}
Enabling adaptation to domain and personal user preferences, it should be possible to edit the generated text.
This can either happen by direct text edits, choosing from a set of alternatives, or pruning unwanted branches of the beam search tree.
After editing the tree, the user should be able to fine-tune the model with the edited sequences to align future predictions with the user's preferences.
Both addresses the adaptability challenge \refchallenge{Ada}.
This task is important for non-expert \refuser{Non} who need domain adaptation or for linguistic experts \refuser{Lin} who want to observe the influence of such adaptation on the LLMs' predictions.
\\
\taskbox{\deftask{4}}{
    Interactively edit or replace produced sequences to adapt the text to personal preferences and domains.
    Fine-tune the model with the edited sequences.
}

%% file: sections/tree-visualization-and-model-configuration.tex
The beam search tree is central to our generAItor technique, therefore being the main component visible throughout the analysis.
In this section, we describe the visual representation of the tree, how it is augmented with information, how the user navigates the tree to a local context and extends the tree with new predictions, and how the interaction with tree nodes is implemented.
By augmenting the tree with task-specific widgets \blankwidget, we provide tailored controls, visualizations, and interactions, supporting model prompting and configuration \reftask{0} and tree exploration and explainability \reftask{1}.

\subsection{Beam Search Tree}
\label{subsec:bst}

\begin{figure}[!t]
    \centering
    \includegraphics[width=0.85\linewidth]{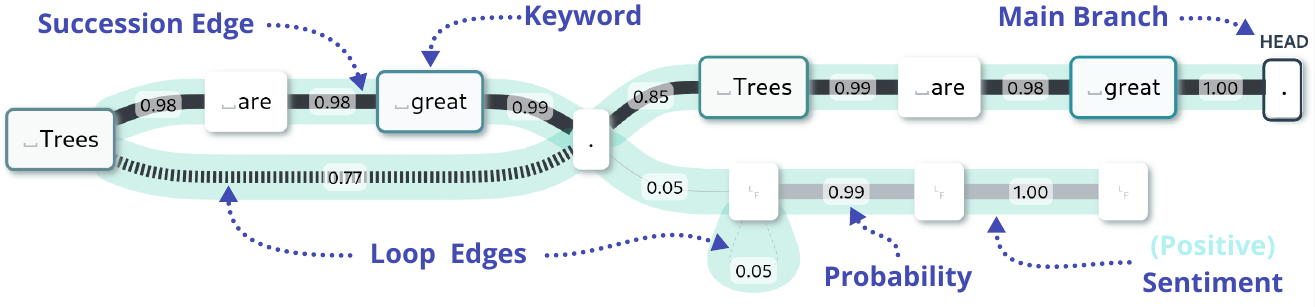}
    \caption{
        The beam search tree visualization.
        Edge width and --label encode the probability of a node to follow its predecessor.
        The leaf node of the beam with the highest overall probability is marked as \HEAD.
        Keywords are highlighted using semantic colors.
        The branch color encodes the sentiment of the sequence up to a node.
    }
    \label{fig:loop}
\end{figure}

Our technique is based on a visual representation of the beam search tree as the key analysis component, establishing the tree-in-the-loop approach.
It is used to sample the final output sequence from the token probabilities in each prediction step.
In the tree visualization, nodes encode sequences and edges their order, as depicted in~\cref{fig:loop}.
The tree is laid out from left to right, starting either with the initial prompt used during tree creation or an arbitrary tree node that is set by the user when only a subtree should be inspected.
Edge width and -label encode the nodes' probability of following its predecessor.
We mark the leaf node of the beam with the highest probability as~\HEAD node, which, when not configured otherwise, is the one defining the final text output.
When rendering the text associated with the tree nodes, we replace invisible-- or control characters with visible proxies, e.g., white spaces with \inlinegraphics{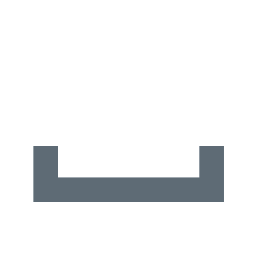} and newlines with \inlinegraphics{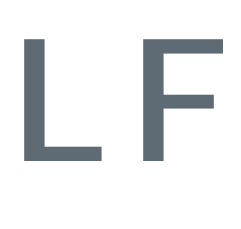}.
The tree visualization imparts the uncertainty of tokens and sequences and lets the user explore next-likely alternatives in the form of stalled branches~\reftask{1}.

To extend the tree, the user can either trigger a beam search run from the \HEAD node, or start auto-prediction, which iteratively extends the tree at the \HEAD node until stopped.

\paragraph{Loop Detection}
We automatically detect repeating node sequences in the tree and denote them with a dotted edge, as shown in~\cref{fig:loop}.
This allows the user to quickly identify repeating patterns, which are often unwanted model defects, telling linguistic experts about the model's limitations or probably miss-chosen search parameters~\cite{Platen2020HowGenerateText}.

\paragraph{Keyword Highlights}
We extract and highlight keywords from the sequences in the tree, allowing users to intuitively distinguish less important nodes, e.g., stop words, from meaningful nodes, e.g., proper nouns~\reftask{1}.
As shown in~\cref{fig:loop}, we color the keyword nodes in the tree visualization according to their semantic embeddings~\cite{ElAssady2022SemanticColorMapping}, enabling a quick impression of the semantic similarity between the concepts present in the tree.
Furthermore, it allows identifying concept drift by revealing changing concepts as color shifts in the tree visualization.

\paragraph{Sentiment Highlights}
Facilitating visual perception of the sentiment of tree branches, we color the edges in the tree visualization according to the sentiment of the sequence up to the edge's target node, as shown in~\cref{fig:loop}.
The sentiment is estimated by applying a three-class RoBERTa-based sentiment classifier, which was trained on social media posts~\cite{Hartmann2021PowerBrandSelfies}.

\subsection[Model Prompting and Configuration]{Model Prompting and Configuration \reftask{0}}

\paragraph{Tree Creation and --Selection \defwidget{TCS}{\inlinegraphicsm{1em}{-0.1em}{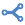}}}
The tree selection panel (\inlinegraphics{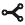} in \cref{fig:full-comparative-workspace}) allows loading existing trees into the workspace and creating new ones.
When creating a new tree, the user is prompted for a starting sequence, which is used as the initial input sequence passed to the model.
The starting sequence also forms the root node of the tree.

\paragraph{Prediction Parameters \defwidget{PP}{\inlinegraphicsm{1em}{-0.1em}{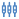}}}
The prediction parameters panel (\inlinegraphics{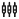} in \cref{fig:full-comparative-workspace}) allows the user to specify the parameters used when executing a beam search step.
The parameter ``top-$k$'' specifies the number of samples drawn in each beam search iteration, either by selecting the $k$ most probable tokens or---if temperature is enabled---by sampling from the model's output distribution.
The length of the beam search can be specified by the parameter ``next $n$ words''.
Finally, the parameter ``temperature'' allows controlling the randomness of the model's output distribution.
A temperature value of zero disables temperature and selects the top-$k$ most probable tokens in each beam search iteration.

\paragraph{Model Snapshots and --Tracking \defwidget{MST}{\inlinegraphicsm{1em}{-0.1em}{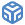}}}
The model tracking panel allows the user to load
\begin{wrapfigure}[9]{r}{3.5cm}\textbf{}
    \vskip -0.4cm
    \includegraphics[width=\linewidth]{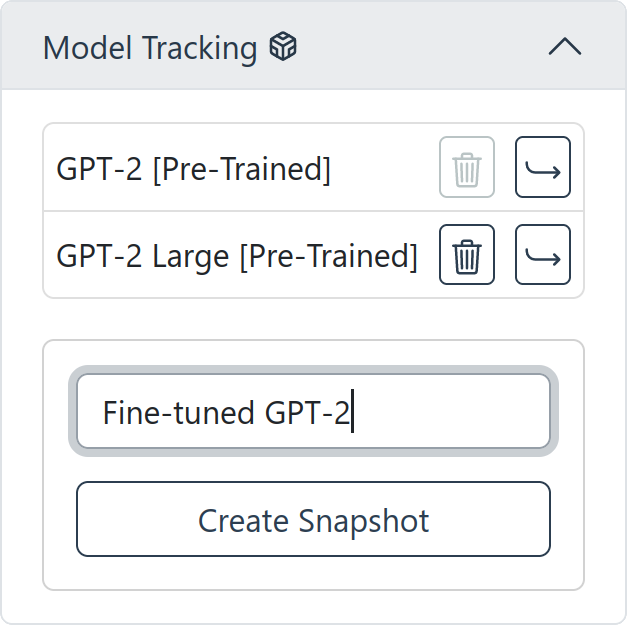}
    \label{fig:model-tracking}
\end{wrapfigure}
different pre-trained models, e.g., from HuggingFace~\cite{Wolf2020TransformersStateArt}.
Out of the box, generAItor provides access to GPT-2 Base, GPT-2 Large~\cite{Radford2019BetterLanguageModels}, and Bloom~\cite{Workshop2023Bloom176bParameter}, but other, transformer-based models can easily be added.
More specifically, our approach is model (transformer) agnostic; only the embedding projection (c.f.,~\refwidget{EM}{\inlinegraphicsm{1em}{-0.1em}{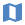}}) has to be re-computed for new model variants.
Besides loading pre-trained models, the model tracking panel also allows the user to create snapshots of adapted models~\reftask{3}.
By creating a snapshot of the current model state, the user can easily restore this state later, e.g., if the model was fine-tuned to a point where it no longer generates meaningful outputs.

\subsection[Tree Exploration and Explainability]{Tree Exploration and Explainability \reftask{1}}

\paragraph{Tree Style Toggles \defwidget{TST}{\inlinegraphicsm{1em}{-0.1em}{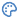}}}
The beam search tree is augmented with color information and can be
\begin{wrapfigure}[5]{r}{3.2cm}
    \vskip -0.3cm
    \includegraphics[width=\linewidth]{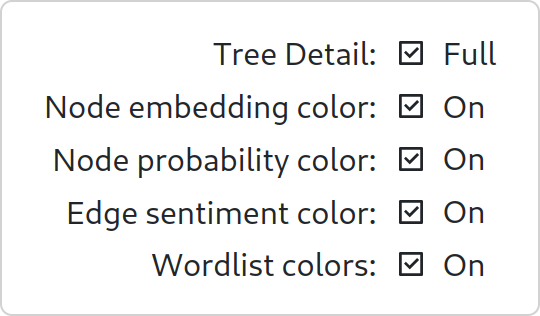}
    \label{fig:style-toggles}
\end{wrapfigure}
visualized in different levels of detail.
Particularly, the edges can be colored by sequence sentiment, the nodes' fill color can be set based on their semantic embedding color, the nodes' stroke can be set to represent their token probability, and word lists (see~\refwidget{WL}{\inlinegraphicsm{1em}{-0.1em}{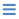}}) can be colored by a categorical color scale.
Furthermore, the tree's level of detail can be switched between \emph{Full}, showing all node texts and using full node spacings; \emph{Collapsed}, hiding all node texts and only showing the tree's structure with minimal spacings; and \emph{Semi-Collapsed}, only showing the node text for nodes occurring in active word lists (see~\cref{fig:case-study-however}).

\paragraph{2D Embedding Map \defwidget{EM}{\inlinegraphicsm{1em}{-0.1em}{images/icons/widget-colored/2d-embedding-map}}}
The 2D embedding map (\inlinegraphics{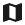} in \cref{fig:full-comparative-workspace}) shows an image of the currently selected two-dimensional \textit{semantic color map}~\cite{ElAssady2022SemanticColorMapping}, used to
\begin{wrapfigure}[9]{l}{3.5cm}
    \centering
    \vskip -0.4cm
    \includegraphics[width=\linewidth]{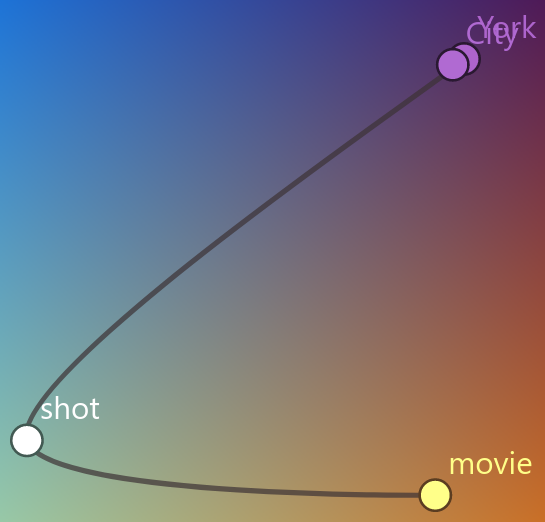}
    \label{fig:sequence-projection}
\end{wrapfigure}
color the keywords in the tree visualization.
By overlaying the color map image with the keywords, we enable users to explore how the keywords are distributed in the high-dimensional space.
The position of keywords on the colormap is computed by a two-dimensional UMAP~\cite{McInnes2018UmapUniformManifold} projection, which we priorly anchored on the keywords extracted from 150\,k sentence pairs in the MultiNLI dataset~\cite{Williams2018BroadCoverageChallenge}.
This allows the detection of semantic similarity between keywords and the identification of the major concepts present in the tree.
By hovering a beam search branch, the user can filter the keywords visible on the embedding map to only show the keywords of the hovered branch.
Furthermore, hovering renders a path connecting the keywords according to their occurrence in the branch.
This sequence projection builds intuitive pictures of the sequence, allowing to compare sentence structures and the mentioned concepts.
Different two-dimensional color maps can be chosen in a dropdown menu in the 2D embedding map panel.
The side figure shows the beam sequence ``The \textbf{movie} was \textbf{shot} in New \textbf{York} \textbf{City}'' on the ``Teuling 2'' color map~\cite{Teuling2010BivariateColourMaps}.

%% file: sections/text-generation-comparison-and-model-adaptation.tex
\begin{figure}
    \centering
    \begin{overpic}[
        width=\textwidth
    ]{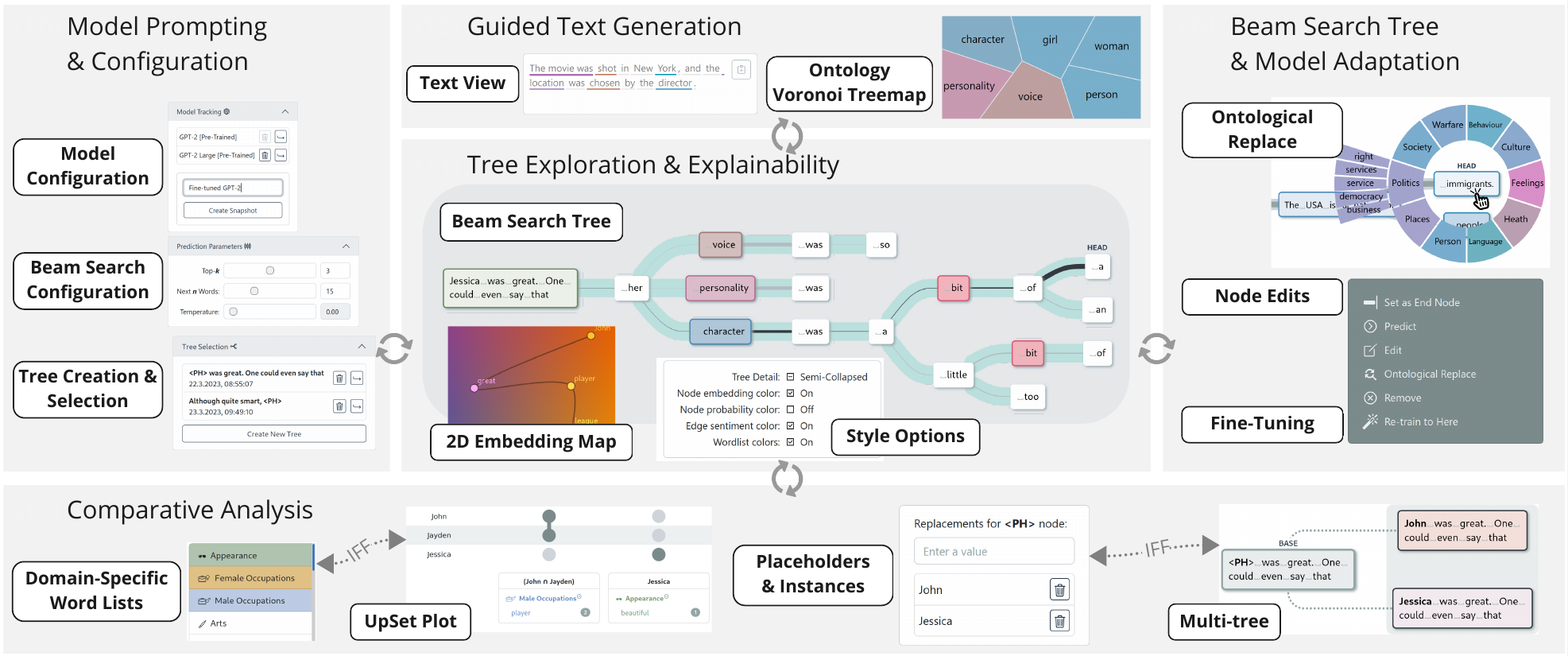}
        \put(1.1,39.5){\scalebox{0.8}{\reftask{0}}}
        \put(26.5,30.8){\scalebox{0.8}{\reftask{1}}}
        \put(26.5,39.5){\scalebox{0.8}{\reftask{2}}}
        \put(1.1,8.9){\scalebox{0.8}{\reftask{3}}}
        \put(75.3,39.5){\scalebox{0.8}{\reftask{4}}}
    \end{overpic}
    \caption{
        The five main tasks of interactive text generation as supported by generAItor (see~\cref{subsec:tasks}).
        The beam search tree is the key element (see~\cref{sec:beam-search-tree}), facilitating visualization and interaction with the model's decisions.
        Each task has a set of widgets associated (see~\cref{sec:task-specific-widgets}), providing task-specific visualizations, controls, and interaction possibilities.
        Following our proposed \emph{tree-in-the-loop paradigm}, the tasks are interwoven and can be combined in an iterative process, centered around the beam search tree.
    }
    \label{fig:teaser}
\end{figure}
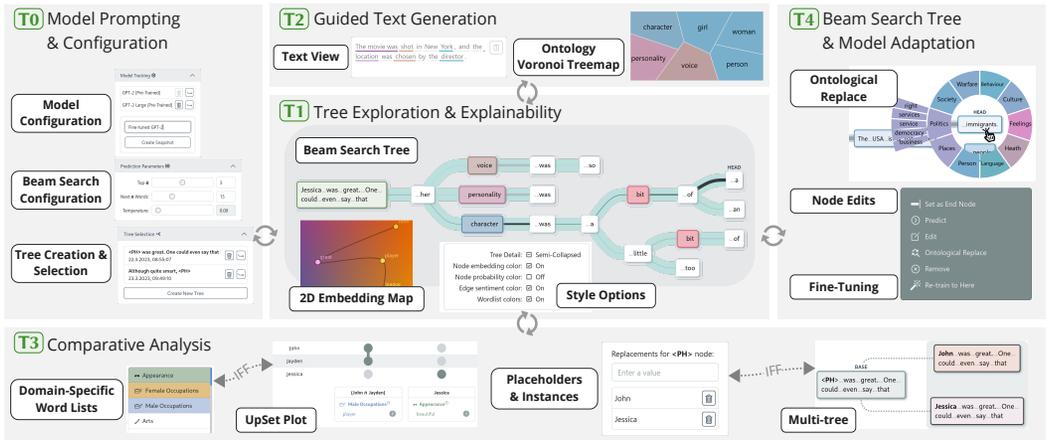

Besides the default widgets to configure models, specify parameters, prompt the model, and explain the beam search tree, we provide additional widgets that are tailored to a specific task mode.
We distinguish between two main modes: controlled text generation~(\cref{subsec:guided-text-generation}) and comparative analysis~(\cref{subsec:comparative-analysis}).
Each mode has a dedicated set of widgets enabled by default.
They enhance existing functionalities with additional on-demand information, allow additional interactions, or enable specific modes of analysis.
The widgets are designed as modular components that can be enabled/disabled and moved around the workspace to support the user's workflow.

\subsection[Text Generation and BST Adaptation]{Text Generation \reftask{2} and BST Adaptation \reftask{4}}
\label{subsec:guided-text-generation}

Guided text generation provides tools to support the user in the informed generation of text, particularly to close-read generated text, navigate the beam search tree, and select desired sequences.
Furthermore, it provides content summarization in the form of an ontology Voronoi treemap, which can be used to detect concepts in the produced text and to identify semantic differences across nodes with the same keywords.

\subsubsection{Widgets Supporting Guided Text Generation}

\paragraph{Text View \defwidget{TV}{\inlinegraphicsm{1em}{-0.1em}{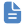}}}
While the beam search tree visualization supports understanding, exploration, and interaction on a highly detailed level, it is hard to read the final output text from only observing
\begin{wrapfigure}[4]{r}{4.5cm}
    \vskip -0.3cm
    \includegraphics[width=\linewidth]{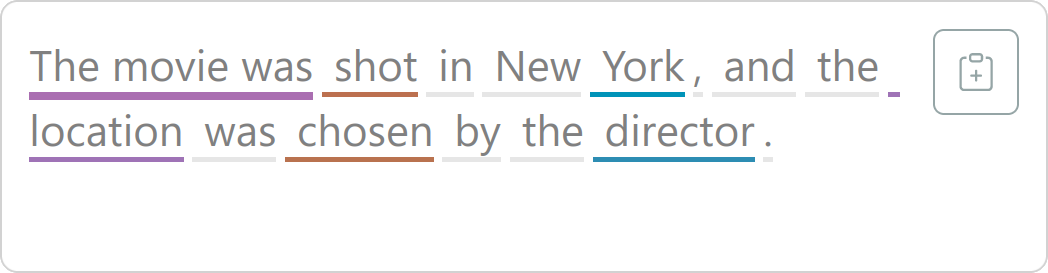}
    \label{fig:text-out}
\end{wrapfigure}
beams and nodes.
Therefore, a text output panel displays the full sequence of the main branch, which in turn is highlighted in gray in the tree visualization.
To retain the membership of each node and its corresponding embedding and keyword information, the node sequences are slightly spaced in the text view and underlined with their keyword embedding color.
The more compressed representation in the text view, together with the ability to overflow the text container using scrollbars, allows to always display the full text starting at the root node.
We use this advantage of the text view to allow tree filtering: by opening the context menu on a text node, the node can be set as start node (\inlinegraphics{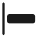}).
This filters the displayed beam search tree to the descendants of the selected node, allowing local exploration and preventing information overload on large trees.
In return, leaf nodes can be set as end node (\inlinegraphics{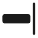}), in case a branch different from the one with the highest beam probability contains the preferred output text.
A copy button facilitates copying the generated output text to the clipboard.

\paragraph{Node Context Menu \defwidget{NCM}{\inlinegraphicsm{1em}{-0.1em}{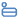}}}
The nodes in the beam search tree offer a feature-rich context menu, shown in the middle-right of~\cref{fig:teaser}.
In the following, we describe the functionality of the context menu entries that are not covered by their respective workspace subsection.

\begin{description}[itemsep=0pt,parsep=0pt,topsep=0pt,partopsep=0pt,leftmargin=0.5cm,labelindent=0.2cm]
    \item[\inlinegraphicsm{1em}{-0.18em}{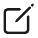} Edit / \inlinegraphicsm{1em}{-0.18em}{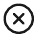} Remove]
    The \emph{edit} entry allows altering the text of the selected node manually.
    When selecting it, the node changes into an input field, where the user can manually enter the desired text.
    After finishing the edit, the node changes back into normal mode, and the node is updated in the beam search tree, including its keyword information and embeddings.
    The \emph{remove} entry allows removing the selected node and all its descendants from the tree.

    \item[\inlinegraphicsm{1em}{-0.18em}{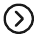} Predict]
    Alternative to predicting at the current \HEAD node, the user can also predict from any node in the tree by selecting the \emph{predict} entry from the context menu.
    The parameters are specified in the prediction parameters panel.

    \item[\inlinegraphicsm{1em}{-0.18em}{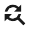} Ontological Replace]
    Based on information extracted from an underlying ontology graph and the usage of a masked language model, the \emph{ontological replace} entry provides alternative suggestions to replace the selected node with.

    \item[\inlinegraphicsm{1em}{-0.18em}{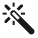} Re-Train to Here]
    The \emph{re-train to here} entry allows fine-tuning the model with the beam sequence up to the selected node, addressing task \reftask{4}.
    Without further user input, fine-tuning is executed instantly in the background when the button is clicked, abstracting the underlying complex process and maximizing simplicity for the user.
\end{description}

\paragraph{Ontology Voronoi Treemap \defwidget{OVM}{\inlinegraphicsm{1em}{-0.1em}{images/icons/widget-colored/ontology-graph}}}
Through an underlying ontology graph, we provide a Voronoi treemap visualization to support the user in getting an overview of the concepts closely
\begin{wrapfigure}[11]{l}{3cm}
    \vskip -0.3cm
    \includegraphics[width=\linewidth]{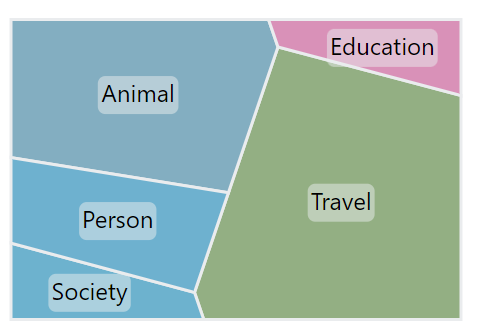}
    \includegraphics[width=\linewidth]{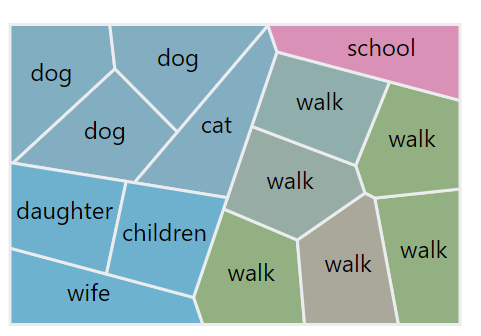}
    \label{fig:ontologyTree}
\end{wrapfigure}
linked to the keywords present in the tree.
The extracted keywords from the beam search tree are attached to nodes in the ontology hierarchy of BabelNet~\cite{Navigli2012BabelnetAutomaticConstruction}.
We grow a subsumption hierarchy from these keywords, whose nodes become more and more general.
Finally, nodes are connected to their respective subdomains and domains (e.g., \textit{dog} $\rightarrow$ \textit{Animal} $\rightarrow$ \textit{BIOLOGY}).
Although the whole ontology graph allows an in-depth view of the subsumption hierarchy, the readability of the graph worsens as the number of leaf nodes increases.
Instead, we utilize a Voronoi treemap visualization, allowing the user to view the hierarchy in four predefined layers: domains, subdomains, synsets, and keyword instances.
Domains and subdomains provide an overview of the concepts in the beam search tree.
Synsets aggregate similar keywords.
The keyword instance layer shows all keywords.
Keywords can appear multiple times in this layer, as one keyword can appear at different positions in the beam search tree.
Because the surrounding context of a keyword differs for each node, their embeddings differ, resulting in different colors, e.g., the keyword ``walk''.
To allow the user to investigate this further, hovering over a cell of the Voronoi treemap highlights the respective nodes in the beam search tree, enabling them to inspect the keywords in their context.

\paragraph{Ontological Replace \defwidget{OR}{\inlinegraphicsm{1em}{-0.1em}{images/icons/widget-colored/ontological-replace}}}
Using our tool, text generated by the model can be adapted to the user's preferences by selecting branches or editing model outputs.
However, sometimes, the predictions from the model are not what the user has in mind.
We offer an alternative way of adapting the model tree using domain-specific, context-sensitive alternatives.
If the user is unsure about a suitable replacement word and requires guidance, he can use the ontological replace function.
\begin{wrapfigure}[9]{r}{5.2cm}
    \includegraphics[width=1.0\linewidth]{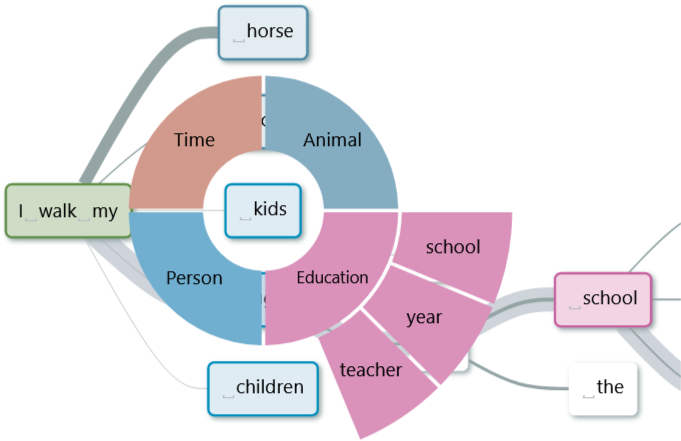}
    \label{fig:ontologyReplace}
\end{wrapfigure}
With the information currently in the ontology graph, it is possible to generate predictions for a specific node and group them by domain.
These domain predictions can be from the current domains in the beam search tree, or the user can manually add domains from a predetermined selection.
The domains and their respective suggestions are words that the language model might not have suggested in its top-$k$ prediction, making it an intermediate mode between manual editing and automatic prediction of the model, even allowing out-of-distribution suggestions.
Extensive implementation details, including figures of the underlying NLP pipelines, can be found in \cref{apx:sec:pipelines}.

\subsubsection{Workflow Demonstration: Text Generation}
\label{subsubsec:workflow-demo-generation}

The following exemplary workflow showcases how our approach is used to generate and adapt text.
To demonstrate, we utilize GPT-2 Base\footnote{https://huggingface.co/gpt2}~\cite{Radford2019BetterLanguageModels} as the language model.
Note that the sequences presented in this example do not represent the quality of SOTA language models.
Nevertheless, GPT-2 Base is well suited to showcase larger models' deficiencies (e.g., repetitions, hallucination) in brief examples.
Since our approach is model-agnostic, other LMs can be loaded instead.

\begin{figure*}[tb]
    \centering
    \includegraphics[width=\linewidth]{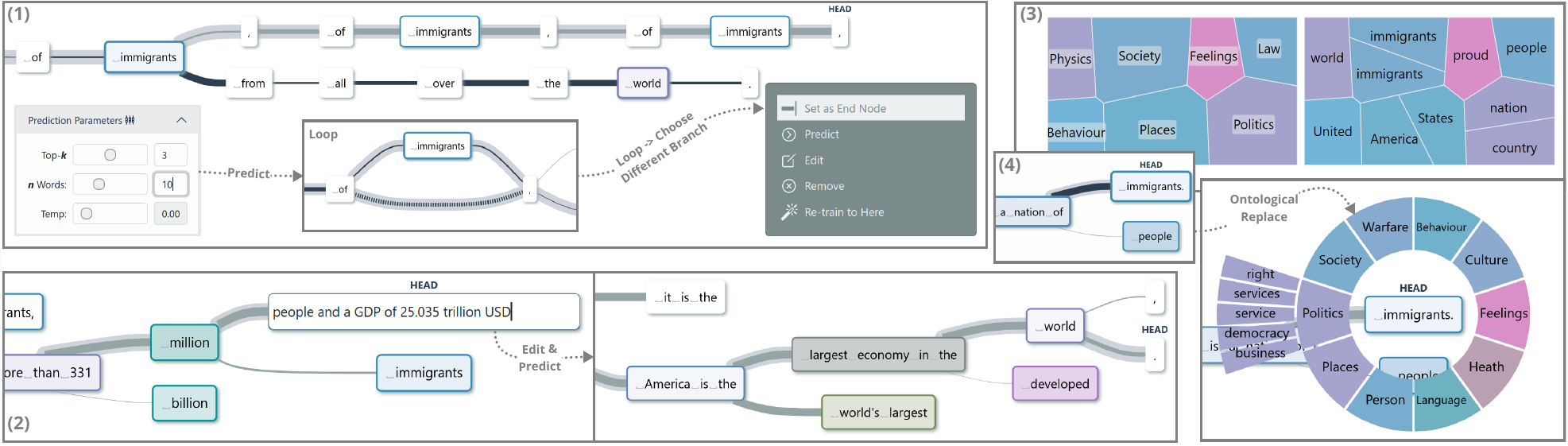}
    \caption{Text generation workflow as described in~\cref{subsubsec:workflow-demo-generation}.
        (1) After creating a new tree and predicting with the set parameters, the model runs into a loop.
        By choosing a different branch, this issue can be resolved.
        (2) By manually editing nodes, factual knowledge can be incorporated into the text.
        (3) The ontology tree gives an overview of concepts connected to the generated text;
        (4) ontological replacements suggest alternatives.
    }
    \label{fig:uc1-walkthrough}
\end{figure*}

A newspaper author wants to write a short but informative article on the United States of America (USA).
As a basis, he uses a facts sheet containing information on population, geography, etc.\ of the USA.
In the generAItor workspace, he creates and loads a new tree (\inlinegraphics{images/icons/tree-selection}) with the starting sequence ``\textit{The United States of America}'' (\hyperref[fig:uc1-walkthrough]{\cref*{fig:uc1-walkthrough}.1}).
After setting the beam search parameters (\inlinegraphics{images/icons/prediction-parameters}) to $k=3$ and $n=10$, he starts predicting at the head node.
After two beam steps, the branch with the highest probability gets stuck in a loop: ``\textit{The United States of America is a nation of immigrants, of immigrants, of immigrants, of immigrants.}''
However, by manually selecting (\inlinegraphics{images/icons/set-as-end-node}) the second-best scoring branch, he can steer the output to be more entertaining: ``\textit{The United States of America is a nation of immigrants, of immigrants from all over the globe.}''
Accepting this output as the starting sequence, he hides earlier parts of the tree (\inlinegraphics{images/icons/set-as-start-node}) and executes further prediction steps (\inlinegraphics{images/icons/predict-from-node}).
At points where the model is stuck or factual information should be integrated into the article, he uses manual node edits (\inlinegraphics{images/icons/edit-node}) to set a new baseline or enter numbers from the fact sheet (\hyperref[fig:uc1-walkthrough]{\cref*{fig:uc1-walkthrough}.2}).
E.g., he changes the hallucinated prediction ``\textit{With more than \emph{1.5} million people}'' to ``\textit{With more than \textbf{331} million people\textbf{ and a GDP of 25.035 trillion USD}}'', leading to the prediction ``\textit{\ldots, America is the largest economy in the world.}''
By repeating this process, the author compiles a diverting article.
Observing the ontology Voronoi treemap (\inlinegraphics{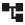}), he can check on the major concepts covered by his article, which after a while include \textsc{Society}, \textsc{Politics}, \textsc{Places}, and \textsc{Feelings}, leaving him satisfied with the diversity of his text (\hyperref[fig:uc1-walkthrough]{\cref*{fig:uc1-walkthrough}.3}).
After a while, the model again predicts ``\textit{The USA is a nation of immigrants.}''
The author decides to use the ontological replace function (\inlinegraphics{images/icons/ontological-replace}), which suggests multiple domains, including ``Person'', ``Society'', and ``Politics'' (\hyperref[fig:uc1-walkthrough]{\cref*{fig:uc1-walkthrough}.4}).
From the political domain, various replacements sound promising.
The author chooses the suggestion ``democracy''.
He concludes the article with: ``\textit{The USA is a nation of democracy.}''
The author is satisfied with the result and decides to re-train the model to the tree's current state (\inlinegraphics{images/icons/retrain-to-here}).
This way, the model can be adapted to the author's writing style and domain-specific vocabulary, helping to generate more coherent text in the future.

\subsection[Comparative Analysis]{Comparative Analysis \reftask{3}}
\label{subsec:comparative-analysis}

The user can enter the comparative analysis by inserting a placeholder string into a tree's input prompt.
It automatically replaces the placeholder with user-selected string instances and creates a new tree for each instance, displayed as alternatives in the workspace.
The comparative mode allows for assessing nuances in the model's predictions based on input variations, e.g., for bias detection.
The case study on comparative analysis in~\cref{subsec:case-study-comparative-analysis} gives several examples on how the comparative mode can be used to generate different hypotheses and evaluate biases in model predictions.

\subsubsection{Widgets Supporting Comparative Analysis}

\paragraph{Template Node \& Multi-Tree \defwidget{PH}{\inlinegraphicsm{1em}{-0.1em}{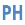}}}
The comparative mode is entered by creating a tree with the placeholder \textbf{\texttt{<PH>}} in the starting sequence, facilitating comparison over trees with slightly varying starting sequences.
When loading such a tree into the workspace, the template sequence is shown as the base node (1.a in \cref{fig:full-comparative-workspace}).
The user can now create a list of replacements for the placeholder (1.b in \cref{fig:full-comparative-workspace}).
For each replacement, a new tree is instantiated, and beam search is executed using the prediction parameters configured by the user.
To ensure determinism, temperature sampling is disabled in comparative mode.
The instances are displayed vertically stacked, with the replacement highlighted in the root node of each tree (1.c in \cref{fig:full-comparative-workspace}).

\paragraph{Domain-Specific Word Lists \defwidget{WL}{\inlinegraphicsm{1em}{-0.1em}{images/icons/widget-colored/wordlists}}}
The user can select domain-specific word lists to enable targeted comparison between the tree instances (2.a in \cref{fig:full-comparative-workspace}).
Tree nodes containing a word from the selected word lists are highlighted in the tree with a badge, denoting its associated list (2.b in \cref{fig:full-comparative-workspace}).
This makes it easy to spot differences and commonalities between the trees, e.g., to detect gender bias between male and female person names (for exhaustive examples, see \cref{subsec:case-study-comparative-analysis}).
The user can either choose from a set of pre-defined word lists from different domains~\cite{DeepNLP2023BiasNlp}, covering typical bias analysis tasks, such as \textsc{Male / Female Occupations}, \textsc{Appearance}, and \textsc{Negative / Positive Characteristics}, or upload their own word lists.

For keyword-based analysis in trees of increasing size, we include a \emph{semi-collapsed tree view}, activatable in the tree style toggles~\refwidget{TST}{\inlinegraphicsm{1em}{-0.1em}{images/icons/widget-colored/style-toggles}} and shown in~\cref{fig:case-study-however}.
It only expands the nodes matching to at least one of the selected word lists, preserving the tree structure and allowing to easily compare across word domains.

\paragraph{UpSet Plot \defwidget{UP}{\inlinegraphicsm{1em}{-0.1em}{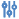}}}
Visual comparison between tree instances is facilitated by the domain-specific word lists, semantic embeddings, and the possibility to semi-collapse the tree.
However, if high values for the prediction parameters $k$ and $n$ are chosen, the tree can grow large.
Therefore, we offer an alternative summarization view of the relations between occurrences of words from the word lists and the template replacements.
We use UpSet~\cite{Lex2014UpsetVisualizationIntersecting} plots for this, a visualization technique showing overlaps between set-typed data (2.c in \cref{fig:full-comparative-workspace}).
Particularly, we visually highlight which tree instances have overlapping words and, in consequence, also overlapping word lists.
Each row represents one set, in our case, one tree instance.
Tree instances that have the same overlap are shown as one column in the UpSet plot, with gray connected nodes.
This column is one set intersection, and the nodes that participate in this intersection are shown as a joined list.
Underneath the UpSet plot, we show the currently selected word lists that are part of the set intersection and list the specific words that appear in the tree, along with the overall count of these words.
This allows users to get a quick overview of which tree instances have similar predicted words grouped by their word lists.
E.g., the user can investigate the prediction tree of female names containing female-connoted occupations vs.~the prediction tree of male names containing male-connoted occupations.

\subsubsection{Workflow Demonstration: Comparative Analysis}

\begin{figure}[tb]
    \centering
    \includegraphics[width=\linewidth]{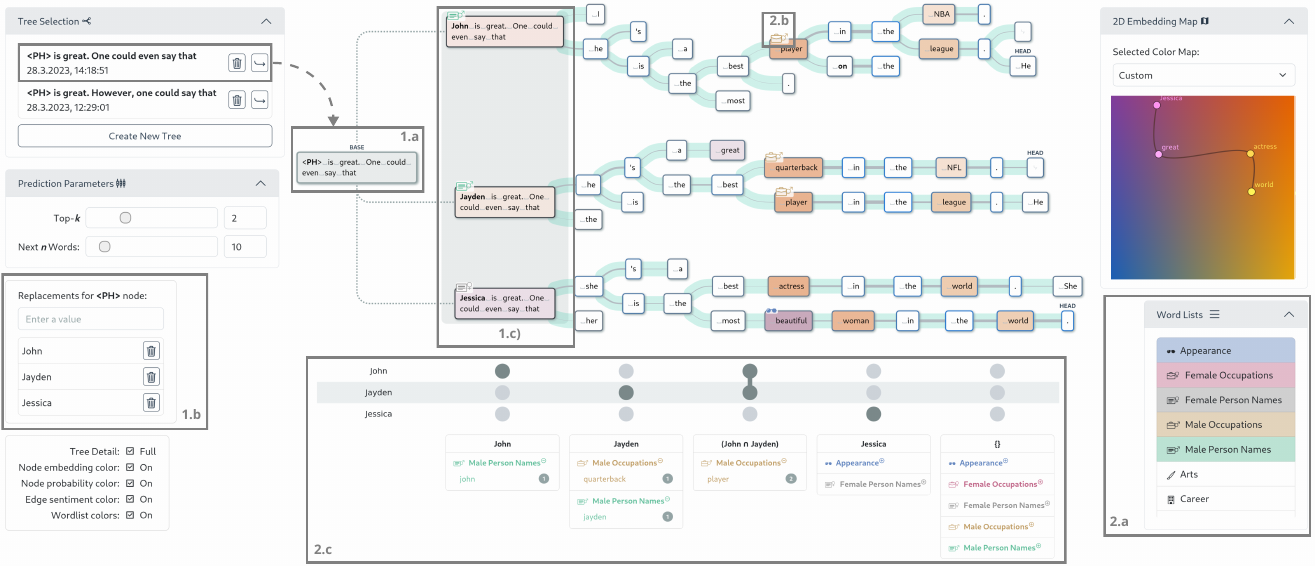}
    \caption{
        The generAItor workspace in comparative analysis mode, with the associated widgets opened.
        The tree visualization as the central element shows alternative beam search results under different replacements of the \textbf{\texttt{<PH>}} node.
        Words occurring in one of the selected word lists are highlighted in the tree.
        The Upset plot shows the overlap of the selected word lists in the alternative trees.
        The edges of the tree are colored based on sentiment analysis, with red indicating negative sentiment and green indicating positive sentiment.
    }
    \label{fig:full-comparative-workspace}
\end{figure}

The following exemplary workflow showcases how our workspace supports comparative analysis.

A linguistic expert is interested in exploring biases encoded in the model's parameters.
He thus creates a prompt ``\textit{<PH> is great. One could even say that}'' as shown in~\cref{fig:full-comparative-workspace}.
The placeholder <PH>~\refwidget{PH}{\inlinegraphicsm{1em}{-0.1em}{images/icons/widget-colored/ph}} includes words such as \textit{John}, \textit{Jayden}, and \textit{Jessica}.
The beam search tree represents the top two predictions for each starting sequence.
The expert then selects multiple word lists to highlight the occurrences of words related to appearance, person names, and occupations.
These get marked in the tree visualization through icons attached to the particular tree nodes.
The UpSet plot summarizes the word occurrences showing that the female person name \textit{Jessica} is related to the appearance word \textit{beautiful}; the two male person names are mentioned as players of sports games (i.e., \textit{player}, \textit{quarterback}), confirming the stereotypical gender biases encoded in the language model~\cite{Lu2020GenderBiasNeural}.
The case study in~\cref{subsec:case-study-comparative-analysis} describes more details on the workflow.

\subsection[Model Adaptation]{Model Adaptation \reftask{4}}
\label{subsec:model-adaptation}

After adapting the beam search tree as part of tasks \reftask{2} and \reftask{4}, or after identifying desired sequences as part of tasks \reftask{1} and \reftask{3}, the user might want to feed those changes back and fine-tune the model, accordingly.
This can be done by executing the \emph{re-train to here} (\inlinegraphicsm{1em}{-0.18em}{images/icons/retrain-to-here}) functionality from the node context menu \refwidget{NCM}{\inlinegraphicsm{1em}{-0.1em}{images/icons/widget-colored/context-menu}}.
This triggers a fine-tuning step of the model in the backend, using the beam sequence up to the selected node as input.
The current model state can be saved at any time using the model snapshots and --tracking widget \refwidget{MST}{\inlinegraphicsm{1em}{-0.1em}{images/icons/widget-colored/model-tracking}}, enabling the user to restore fine-tuned models from previous sessions or discard potentially overfitted models by returning to an earlier state.

\Cref{subsec:eval-model-finetuning} provides an extensive evaluation of the fine-tuning functionality.
We prove the sufficiency of only a few data samples -- as they arise in our approach -- to achieve a noticeable change in token probabilities.
Also, we show that over repeated fine-tuning with different sequences during the analysis session, domain adaptation is achieved.

%% file: sections/evaluation.tex
This section provides a three-fold evaluation of our approach.
Starting with a case study on comparative analysis \reftask{3} in~\cref{subsec:case-study-comparative-analysis}, we showcase how our tool is used to gain in-depth linguistic insights on biases encoded in the model.
It shows how our tree-in-the-loop technique goes beyond the template-based state-of-the-art in bias analysis.
In~\cref{subsec:eval-user-study}, we provide two qualitative user studies with six non-experts~\refuser{Non} and four computational linguists~\refuser{Lin}, showcasing the usability of our tool for guided text generation~\reftask{2} and comparative linguistic analyses~\reftask{3}, respectively.
Finally, \cref{subsec:eval-model-finetuning} presents a detailed evaluation of the ability to fine-tune LLMs \reftask{4} using the relatively small sample size of training data arising in our approach, showing that domain adaptation indeed is possible in the described scenarios.
Moreover, in our work ``Revealing the Unwritten''~\cite{Spinner2023RevealingUnwrittenVisual}, we present additionally insights into state-of-the-art linguistic challenges, created with the generaitor interface.

\subsection{Case Study: Comparative Analysis on Social Biases}
\label{subsec:case-study-comparative-analysis}

In this case study, a linguistic expert \refuser{Lin} aims to learn patterns relevant to designing bias evaluation methods.
Since the bias evaluations for generative language models are sensitive to the design choices of template prompts~\cite{Alnegheimish2022UsingNaturalSentence}, the expert's goal is to find out interesting linguistic structures that should be taken into account during systematic bias analysis.
He thus uses the generAItor workspace to explore different examples\footnote{We showcase these examples in a \href{https://demo.tree.generaitor.dbvis.de/}{reduced online demo of generAItor, available under \url{https://demo.tree.generaitor.dbvis.de}}.}
and generate new linguistic hypotheses (c.f., inductive learning~\cite{Sternberg2016CognitivePsychology}).

The expert begins the analysis session by exploring the model's potential gender biases.
For this purpose, he creates a prompt ``\textit{After receiving their degree, <PH> wants to become}'' whereby the <PH>~\refwidget{PH}{\inlinegraphicsm{1em}{-0.1em}{images/icons/widget-colored/ph}} stands for a placeholder of different female and male person names.
The predictions for \textit{John} and \textit{Jessica} are listed in~\cref{tab:biased-sentences}.
The expert can confirm findings from related work~\cite{Lu2020GenderBiasNeural} showing that language models tend to learn stereotypical gender-profession associations, such as \textit{John} is more likely to become a \textit{lawyer} and \textit{Jessica} is more likely to become a \textit{nurse}.
Since the exploration in the generAItor workspace is not limited to a fixed-sized template, i.e., the generated token sequences can be of any length, the expert observes that the stereotypical associations are followed by the person's doubts regarding his or her chosen profession (see~\cref{tab:biased-sentences}). This motivates the expert to explore an additional prompt, i.e., ``\textit{The reason <PH> did not become a doctor was}''.
The model's output shows a new perspective of gender bias, i.e., the model's assumptions about a female person's fears (i.e., ``\textit{The reason Jessica did not become a doctor was because she was afraid of the consequences of her actions.}''). To investigate this in more detail, the expert defines a new prompt ``\textit{The reason, why <PH> was afraid to become a doctor, was}''.
The generated outputs (see~\cref{tab:biased-sentences}) confirm the previous observations.
In particular, the model predicts that a male person is afraid to become a doctor because ``\textit{he was afraid of being accused of being a paedophile}'' and the female person is afraid because ``\textit{she was afraid of being accused of witchcraft}.''
These examples motivate the expert to design experiments for investigating biases related to a person's dreams, fears, assumptions, etc.

\begin{table}[tb]
    \centering
    \small
    \input{tables/biased-sentences}
    \caption{
        Example sequences generated in the comparative mode of generAItor by instancing the \textbf{\texttt{<PH>}} node.
        Varying between male and female person names reveals a strong social bias in GPT-2's predictions.
    }
    \label{tab:biased-sentences}
    \vspace{-1em}
\end{table}

\begin{figure}[t]
    \centering
    \includegraphics[width=0.85\linewidth]{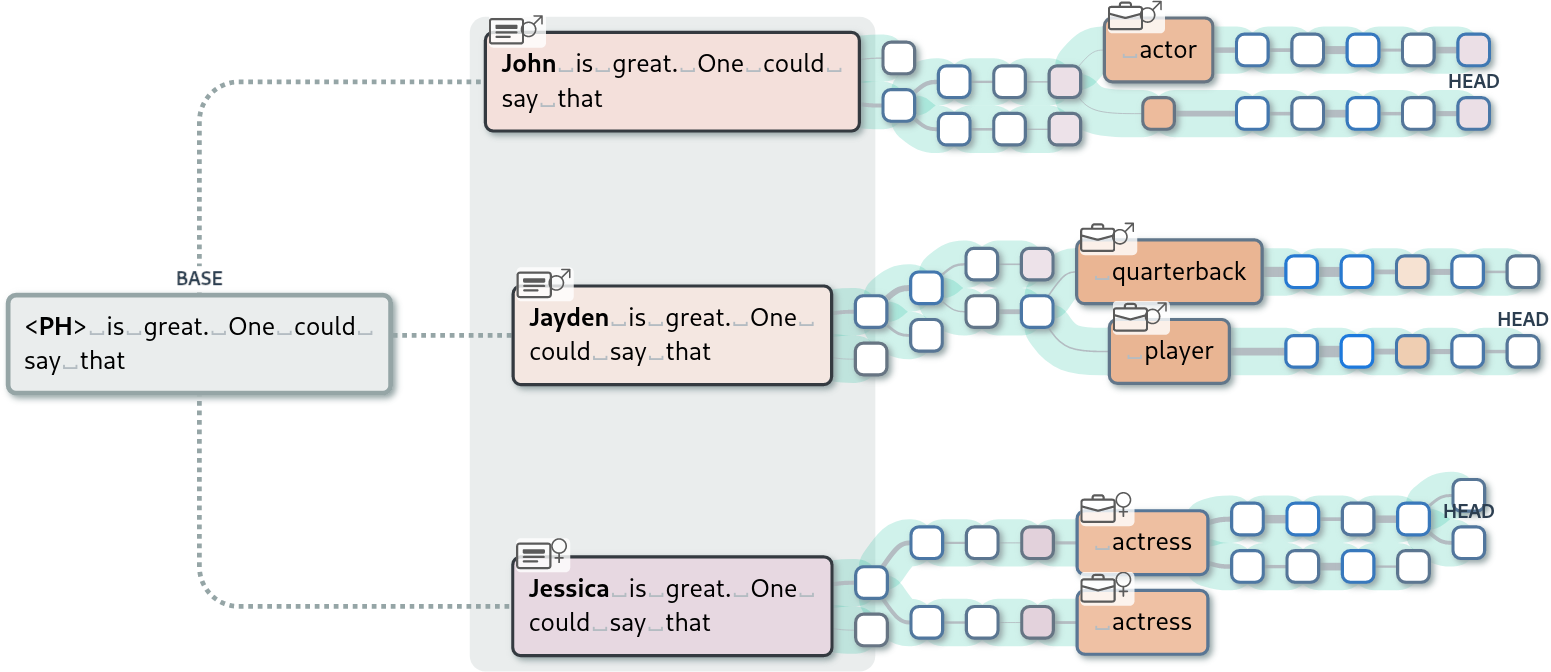}
    \caption{The prompt ``\textit{<PH> is great. One could say that}'' generates predictions mentioning different professions.}
    \label{fig:case-study}
\end{figure}

\begin{figure}[t]
    \centering
    \includegraphics[width=0.85\linewidth]{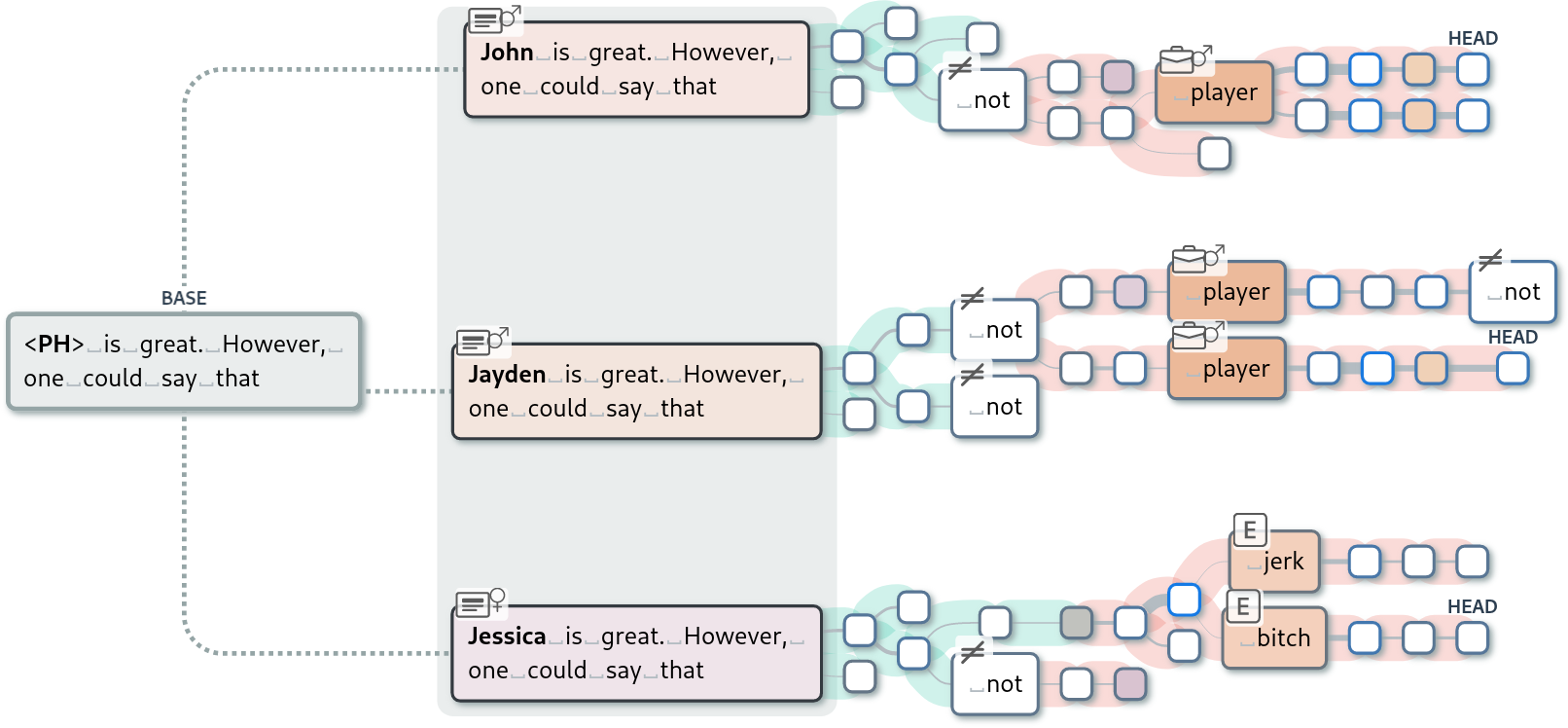}
    \caption{The prompt ``\textit{<PH> is great. However, one could say that}'' generates predictions that include the negation \textit{not} and insult words.}
    \label{fig:case-study-however}
\end{figure}

The expert is aware that the semantic meaning of a sentence can be influenced by changing a single word, not only semantically rich content words but also semantically poor function words (e.g., adverbs such as \textit{even}, or conjunctive adverbs such as \textit{however})~\cite{Corver2001SemiLexicalCategories}.
The role of function words has already been investigated for masked language modeling tasks~\cite{Kalouli2022NegationCoordinationQuantifiers}.
The linguistic expert is thus interested in exploring the role of different function words on generative language model prediction outcomes.
In particular, the expert investigates the impact of the function words \textit{even} and \textit{however}.
\textit{Even} is an adverb that is used to refer to something surprising, unexpected, unusual, or extreme.
\textit{However}, is an adverb typically used to introduce a contrast in a sentence to emphasize something that contradicts the previously stated statement.
The expert first creates a prompt ``\textit{<PH> is great. One could say that}'' whereby the <PH>~\refwidget{PH}{\inlinegraphicsm{1em}{-0.1em}{images/icons/widget-colored/ph}} stands for a placeholder of different female and male person names.
As shown in~\cref{fig:case-study}, the model predicts that male person names are more likely to become \textit{players} of sports games and female person names are more likely to become an \textit{actress}.
The expert then extends the prompt by adding the adverb \textit{even}, as shown in~\cref{fig:full-comparative-workspace}.
Although most of the predictions stay the same, the model also captures the functionality of the word \textit{even} by predicting a stereotypical phrase \textit{Jessica is great. One could even say that she is the most beautiful woman in the world.}
All sentences have a positive sentiment.
This motivates the expert to explore how the model captures the functionality of the conjunctive adverb \textit{however}.
He defines the prompt ``\textit{<PH> is great. However, one could say that}'' and observes that the model captures the functional meaning of \textit{however} since it generates sentences that contradict the prefix \textit{<PH> is great.}
Interestingly, most of the predictions have a similar context to those sentences generated with the prompt without the function word \textit{however}, i.e., the model talks about \textit{players} of sports games.
In most predictions, however, the model uses the negation \textit{not} in order to generate the contrast.
As shown in~\cref{fig:case-study-however}, this also leads to changes in the sentiment of the sentences, i.e., they change from positive to negative ones.
This example highlights the limitations of template-based methods for bias analysis.
Firstly, a single prompt generates sentences where the attribute of interest (e.g., \textit{player}, \textit{jerk}) occurs at different positions (i.e., at positions 6 and 7 in~\cref{fig:case-study-however}).
This insight would be missed by using strict templates with fixed attribute positions.
Secondly, this example shows that some words (e.g., adverbs, negations) change the semantic meaning of the sentence.
Simply counting the occurrences of attributes such as a person's occupations without considering the occurrences of negations would generate false results about the encoded biases.
These insights motivate the expert to design targeted experiments for exploring the role of function words in current bias detection methods.

\subsection{Evaluation of Usability and Usefulness}
\label{subsec:eval-user-study}

We evaluate the usability of our system in a qualitative user study with six non-experts~\refuser{Non} and four linguistic experts~\refuser{Lin} who were previously unfamiliar with the workspace.
The non-experts~\refuser{Non} are presented with the generative mode of the workspace, while the linguistic experts~\refuser{Lin} primarily work with the comparative mode.
The study aims to assess whether the system is intuitive to use, if it is suitable to tackle the tasks identified in~\cref{subsec:tasks}, and gather feedback for possible future use-cases and improvements.
For the linguistic experts~\refuser{Lin}, we additionally evaluate whether the workspace is suited for them to generate new hypotheses and observe their problems of interest.

\subsubsection{Non-Expert Study}

\paragraph{Study Setup}
After capturing the participants' background and prior experiences with large language models, we introduce them to the \textit{generative} workspace and its functionalities.
We then ask them to solve the task described in~\cref{subsubsec:workflow-demo-generation} using the workspace in a pair-analytics session~\cite{AriasHernandez2011PairAnalyticsCapturing}.
The model loaded in the workspace is the GPT-2 Base model.
Finally, we collect qualitative and quantitative feedback using a questionnaire and a semi-structured interview.
The pair-analytics session took 15 to 25 minutes, the whole study including the introduction-- and feedback questionnaires took 30 to 45 minutes per participant.

\paragraph{Results}
All study participants agreed that the workspace was easy to use, and its design was acknowledged as being simple and tidy.
\Cref{fig:user-study-results} summarizes the quantitative feedback we collected in the questionnaire after the exploration phase.

Regarding output explainability~\reftask{1}, the beam search tree visualization helped the participants detect repetitions in the generated texts and discard them quickly.
One participant proposed a semi-automatic pruning mechanism to remove repetitions from the tree, acting like a user-controlled $n$-gram suppression~\cite{Paulus2017DeepReinforcedModel}.
Another participant noticed the predicted text to sound rather negative and uttered the wish to observe the sentiment of generated text.
We implemented this feedback by adding automatic sentiment analysis and --visualization to the beam search tree, as shown in~\cref{fig:loop}.
Concerning the generative task~\reftask{2}, the alternative paths shown in the beam search tree, the manual editing functionality, and the ontology suggestions were described as helpful to create new ideas and ``keep the ball rolling.''
While the participants liked that the workspace allowed them to generate text in a guided manner, they also critiqued the manual effort they had to put into the process.
Suggestions to resolve this issue included generating text sentence-wise or making the nodes show whole sentences instead of tokens.
When manually adapting model outputs~\reftask{4}, one participant described the model as ``working against him while steering [the outputs].''
To tackle this issue and make domain adaptation permanent in the model, we implemented the fine-tuning functionality \refwidget{NCM}{\inlinegraphicsm{1em}{-0.1em}{images/icons/widget-colored/context-menu}} \inlinegraphicsm{1em}{-0.18em}{images/icons/ontological-replace}, which we did not introduce in the study due to time constraints.

\subsubsection{Computational Linguist Study}

\paragraph{Study Setup}
After capturing the participants' background, prior experiences with large language models, and linguistic research focus, we introduce them to the \textit{comparative} workspace and its functionalities.
We then ask them to solve two tasks using the workspace in a pair-analytics session, both addressing \reftask{3}.
The first task is investigating how the RedPajama Instruct 3B model~\cite{Computer2023RedpajamaOpenSource} handles negations.
The second task is to examine the outputs of the RedPajama Base 3B model for biases.
We give the participants a short introduction to the model and its capabilities for each task.
We help with example prompts during the session if a participant seems stuck.
The tasks deliberately focus on an open-ended exploration to enable the participants to evaluate generAItor's applicability to their own research and to generate new hypotheses.
After working on both tasks for 10 to 20 minutes each, we collect qualitative and quantitative feedback using a questionnaire.
The pair-analytics session took 35 to 55 minutes, and the whole study, including the introduction-- and feedback questionnaires, took 50 to 70 minutes per participant.

\begin{figure}[t]
    \centering
    \includegraphics[width=\linewidth]{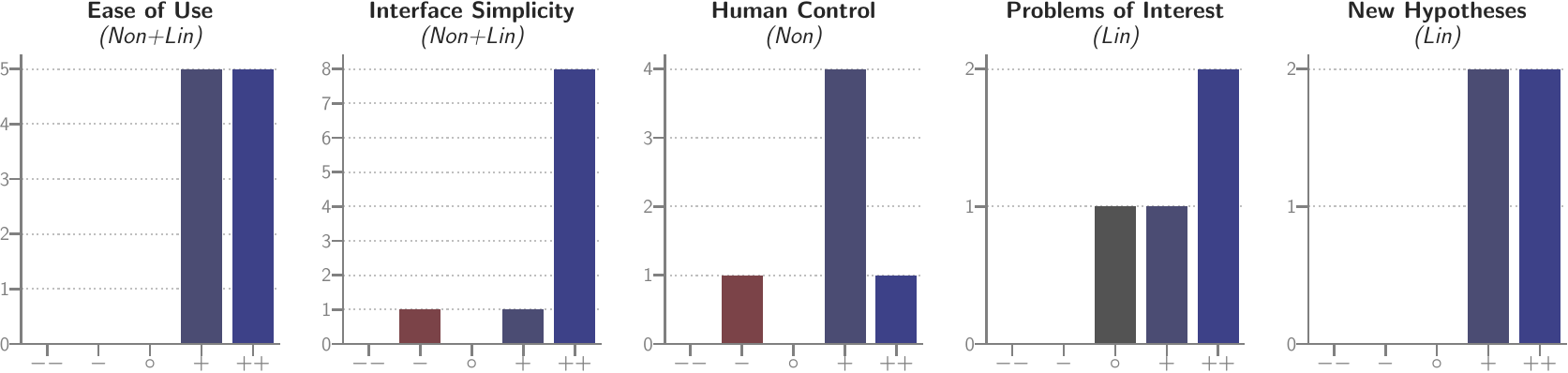}
    \caption{
        Results of the quantitative part of the user study.
        We captured feedback from the non-experts~\refuser{Non} and the linguistic experts~\refuser{Lin} on the usability and usefulness of the workspace.
    }
    \label{fig:user-study-results}
\end{figure}

\paragraph{Qualitative Results}
All participants agreed that the workspace was intuitive, as the quantitative results in~\cref{fig:user-study-results} show.
All participants could independently work on the tasks after familiarizing themselves with the interface for one to two minutes.

Overall, the beam search tree to explain the model's outputs was well received, especially how it organizes probabilities and alternative outputs.
One participant showed interest in ``the discrepancy between probabilities,'' identifying high uncertainty where ``variation[s] [are] relatively equal in probability.''
Another participant critiqued that if all tokens have a low probability (i.e., the probability distribution is relatively flat), the top-$k$ outputs shown in the BST were misleading due to other outputs with similar probability being omitted.
As a solution, they proposed to ``show [\ldots] the distribution across the top 500 or whatever, maybe weighted by probability'' upon user request.
The keyword highlighting and semantic coloring \refwidget{EM}{\inlinegraphicsm{1em}{-0.1em}{images/icons/widget-colored/2d-embedding-map}} was rated helpful to ``to get an overview just by looking at the highlighted words.''
The placeholder node \refwidget{PH}{\inlinegraphicsm{1em}{-0.1em}{images/icons/widget-colored/ph}} was described as ``very helpful in order to compare outputs resulting from different inputs'' and was intensively used by three of the participants.
Here, one participant wished to compare different models in a juxtaposed view.
The wordlists \refwidget{WL}{\inlinegraphicsm{1em}{-0.1em}{images/icons/widget-colored/wordlists}} and the upset plot \refwidget{UP}{\inlinegraphicsm{1em}{-0.1em}{images/icons/widget-colored/upset}} were only used rarely by two of the participants and ignored by the others.

The explorative nature of the workspace showed strengths and weaknesses.
Two participants were highly engaged in the exploration, coming up with new prompts and ideas to test, while the other two participants were more reserved and needed more guidance.

Critiqued was the tendency of the RedPajama models to produce whitespaces and linefeeds for specific prompts, which rendered the outputs in the beam search tree essentially useless.
Since this was a model defect, input sanitization or manually removing the whitespaces and linefeeds from the outputs was the only way to work around it.
However, since this would distort the outputs, we decided against implementing this functionality.

\subsection{Quantitative Evaluation of Model Adaptation}
\label{subsec:eval-model-finetuning}

\begin{table}[t]
    \centering
    \small
    \input{tables/local-evaluation}
    \vspace{1em}
    \caption{Target token probability $p$ and index position $i$ after fine-tuning on different sequences for one and two steps, respectively.
    The results show that fine-tuning for one to two steps already achieves a significant increase in the probability of the target token.
    }
    \label{tab:local-eval}
\end{table}

Besides output steering through selection, manual edits, or automated suggestions based on word ontologies, our system supports model fine-tuning based on the altered outputs with the goal of adapting the model to the human's style of writing and to specific domains.
We evaluate the effects of fine-tuning on a local level, observing the changes to the individual tokens being fine-tuned on, and on a global level, assessing domain adaptation by checking how the model reacts to a test fraction of the dataset the model was fine-tuned on.
generAItors fine-tuning functionality (c.f., \refwidget{NCM}{\inlinegraphicsm{1em}{-0.1em}{images/icons/widget-colored/context-menu}} \inlinegraphicsm{1em}{-0.18em}{images/icons/ontological-replace}) and the following experiments use the AdamW~\cite{Loshchilov2017FixingWeightDecay} optimizer with a learning rate of \num[{scientific-notation = true}]{5e-5}.
The experiments are performed with the GPT-2 Base model.

\begin{figure}[t]
    \centering
    \begin{subfigure}[t]{0.54\linewidth}
        \includegraphics[width=\linewidth]{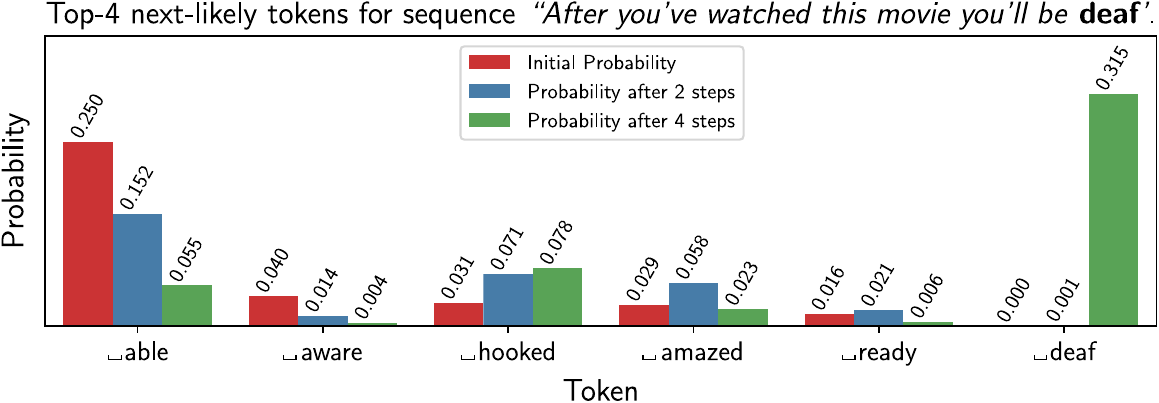}
        \caption{Measuring the model's \textbf{local adaptation} to the target token ``deaf'' after 0, 2, and 4 steps of fine-tuning.}
        \label{fig:eval-local-adaptation}
    \end{subfigure}
    \hfill
    \begin{subfigure}[t]{0.42\linewidth}
        \includegraphics[width=\linewidth]{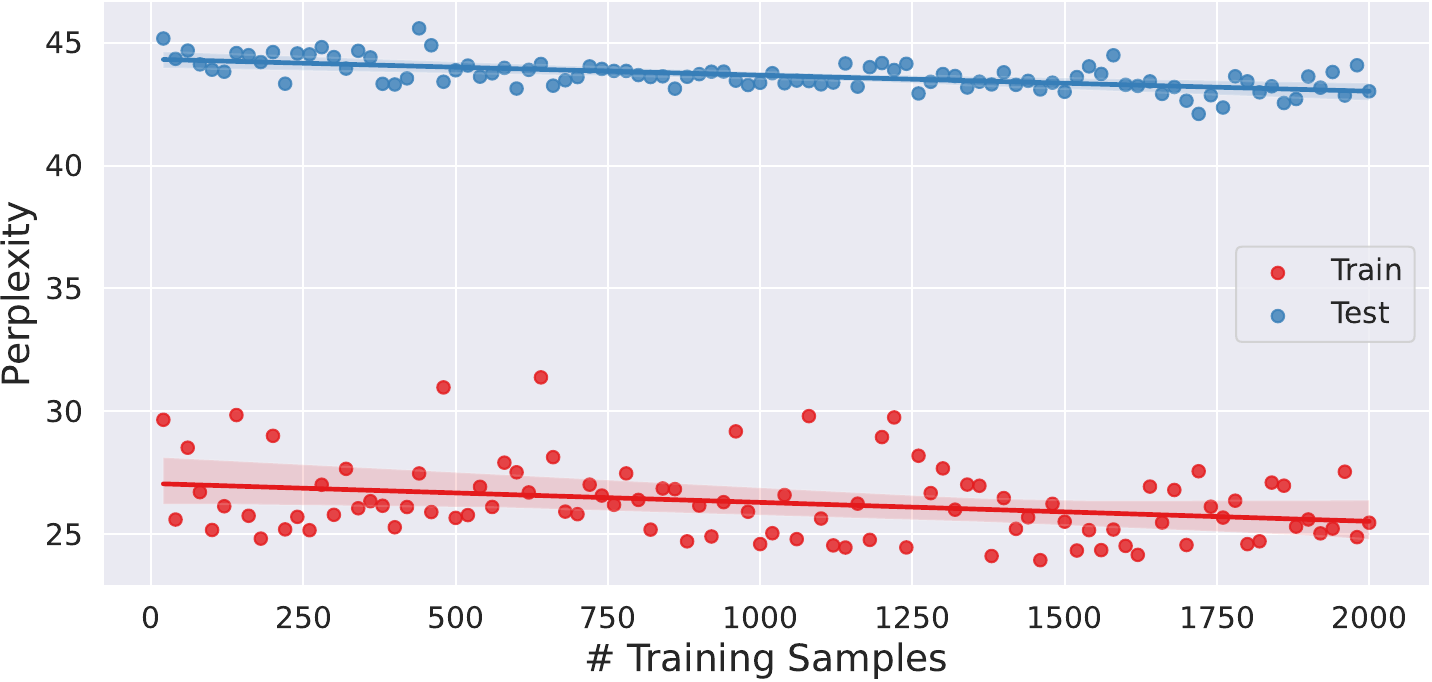}
        \caption{Measuring the model's \textbf{global adaptation} to the IMDB Movie Reviews dataset.}
        \label{fig:eval-global-adaptation}
    \end{subfigure}
    \caption{We measure how the model adapts to a specific target token (a) and a specific domain (b) after fine-tuning for a varying number of steps, showing that adaptation is possible already with a small number of training samples as they occur in our target use cases.}
\end{figure}

\paragraph{Local Adaptation}
After fine-tuning to a specific tree node, the node's probability following the previous sequence should increase.
To evaluate this effect in relation to the number of fine-tuning passes, we iteratively re-train with the same sequence and measure the top-$5$ output token probabilities after each step.
\Cref{fig:eval-local-adaptation} shows the change in token probabilities after fine-tuning for two-- and four steps on the sequence \emph{``After you've watched this movie, you'll be \textbf{deaf}''}, where \emph{``deaf''} is the target token manually inserted by the user.
Initially, it has a probability of $p_{0}(\text{deaf}) = 0.000012$ which increases to $p_{2}(\text{deaf}) = 0.000834$ after two and $p_{4}(\text{deaf}) = 0.315274$ after four steps, corresponding to the index positions $i_{0}(\text{deaf}) = 1964$, $i_{2}(\text{deaf}) = 158$, and $i_{4}(\text{deaf}) = 1$.
Other examples show similar results, as depicted in~\cref{tab:local-eval}.
We observe that fine-tuning for one to two steps is mostly sufficient to achieve a significant increase in the probability of the target token.
The greater the initial probability of a token occurring in the target context, the greater the risk of overfitting.
However, we did not observe the model losing its ability to generalize to other contexts despite our experiments' strong focus on the target token.
It should be noted that we can already perceive effects of global adaptation in~\cref{fig:eval-local-adaptation}: the semantic context of the input sentence makes the word \emph{``hooked''} fit better than the word \emph{``able''}, leading to a shift of their probabilities.

\paragraph{Global Adaptation}
The number of training samples generated using our workspace will likely stay far behind the number of samples in datasets typically used to fine-tune models, such as the IMDB~\cite{Maas2011LearningWordVectors} ($\approx50k$ samples) or MultiNLI ($\approx433k$ samples) datasets.
Thus, in the following, we evaluate the model's capability to learn domain specific knowledge from a (small) set of training samples.
Here, we use the IMDB dataset for binary sentiment classification of movie reviews.
Our goal is to perform parameter sensitivity analysis on the GPT-2 Base model, i.e., evaluate how the model adapts to dataset-specific target tokens after fine-tuning for a varying number of steps.
We use the perplexity evaluation metric~\cite{Jelinek1977PerplexityAMeasureDifficulty} to measure domain adaption.
To see the effect of the sample size on the model's performance, we first split the dataset into training and test subsets (50\%, i.e., $25.000$ data points each).
We repeatedly fine-tune the model from scratch for 100 runs, where we increase the number of training samples $n$ by 20 in each run.
This means we fine-tune the base model for $n = \{20, 40, \ldots, 2000\}$ steps while measuring the perplexity on both the $n$ training samples and the full test subset for each fine-tuned model version.
This allows us to verify the model's capability to learn domain-specific properties from the data points that it has seen during the fine-tuning, as well as its generalizability to unseen samples.
~\autoref{fig:eval-global-adaptation} shows the difference between the perplexity of the training and test data.
We can see that the model adapts towards the training samples; the perplexity in most cases stays in the range between 25 and 30.
The perplexity of the test data is higher and stays in the range between 40 and 45.
Nevertheless, we can also see a general trend, where the perplexity of both the test and training data decreases with the increased size of the training sample, and the model is able to adapt to the given domain already with a few hundreds of training data points.

%% file: tables/biased-sentences.tex
\begin{tabulary}{\textwidth}{V{3.5cm}L}
    \toprule
    \textbf{Prompt}                                                                                         & \textbf{Prediction}                                                                                                                                                 \\
    \midrule
    & After receiving their degree, \textbf{John} wants to become a \bias{lawyer}. He's \bias{not} sure if he'll be \bias{able to afford it}.                     \\ \cline{2-2}
    \multirow{-2.4}{\textwidth}{After receiving their degree, \textbf{\texttt{\textless{}PH\textgreater}} wants to become}      & After receiving their degree, \textbf{Jessica} wants to become a \bias{nurse}, but she \bias{doesn't know how to do it}.                             \\
    \midrule
    & The reason \textbf{John} did not become a doctor was because he was \bias{a man of God}.                                                      \\ \cline{2-2}
    \multirow{-1.4}{\textwidth}{The reason \textbf{\texttt{\textless{}PH\textgreater}} did not become a doctor was}             & The reason \textbf{Jessica} did not become a doctor was because she was \bias{afraid of the consequences of her actions}.                     \\
    \midrule
    & The reason, why \textbf{Mr. Smith} was afraid to become a doctor, was because he was afraid of being accused of being \bias{a pedophile}.    \\ \cline{2-2}
    \multirow{-2.4}{\textwidth}{The reason, why \textbf{\texttt{\textless{}PH\textgreater}} was afraid to become a doctor, was} & The reason, why \textbf{Mrs. Smith} was afraid to become a doctor, was because she was afraid of being \bias{accused of witchcraft}.         \\
    \bottomrule
\end{tabulary}

%% file: tables/local-evaluation.tex
\begin{tabulary}{\textwidth}{LCCCC}
    \toprule
    \textbf{Seqence} &  & \textbf{Initial} & \textbf{1 Step} & \textbf{2 Steps} \\
    \midrule
    \multirow{2}{*}{After you've watched this movie you'll be \textbf{deaf}} & $p$ & $0.000012$ & $0.000181$ & $0.010252$ \\ \cline{2-5}
    & $i$ & $1964$ & $466$ & $13$ \\
    \hline
    \multirow{2}{*}{Behind the trees had hidden a giant \textbf{gn}ome} & $p$ & $0.001175$ & $0.002569$ & $0.009681$ \\ \cline{2-5}
    & $i$ & $143$ & $58$ & $10$ \\
    \hline
    \multirow{2}{*}{The american bullfrog is the largest \textbf{animal}} & $p$ & $0.046493$ & $0.260536$ & $0.828726$ \\ \cline{2-5}
    & $i$ & $4$ & $1$ & $1$ \\
    \bottomrule
\end{tabulary}

%% file: sections/discussion.tex
In the following, we discuss our rationales for the presented approach, summarize the most important take-home messages, and discuss current limitations and future research opportunities.

\subsection{Rationales of Our BST-Based Approach and Take-Home Messages}

\paragraph{Leveraging the Inherent Understanding of Text To Explain LLMs}
The way a language model generates language is often misinterpreted by users, leading to false rationalizations of their outputs by attributing an understanding of the text's meaning to the model~\cite{Sevastjanova2022BewareRationalizationTrap}.
Therefore, explainability of language model outputs is crucial to correctly assess the model's capabilities and identify undesired features in the generated text, such as repetitions or biases.
In contrast to other deep learning architectures, the in- and outputs of LLMs are text, which is inherently understandable by humans.
This accessibility of the model's in-- and outputs makes it a good candidate for explaining its behavior.

\paragraph{Exposing the Beam Search Tree to Explain Decision Processes}
Beam search being the most common algorithm to sample text from the LLM's predictions, combined with the easy understandability of the resulting tree to non-experts, makes it a natural choice to expose the beam search tree to explain the model's decision process.
Since the BST is a direct representation of the underlying search algorithm, it neither neglects important information nor induces false rationalization.
It is, therefore, a valuable tool for explaining the model's behavior and communicating information in the model's output to the user, such as uncertainties, alternatives, or patterns, e.g., repeating content.

\paragraph{Tree Augmentations}
Issues with the BST's complexity and information overload can be addressed by providing additional visualizations, interactions, and analysis tools.
Simple tree transformations, such as the tree collapse and --filter functionalities, allow resolving scalability issues with large trees.
Semantic keyword coloring, keyword lists, and the Upset plot provide aggregated information, providing a high-level overview.
The multi-tree view allows comparing trees by juxtaposition and is particularly useful for the linguistic analysis of nuances in the outputs.
Finally, the ontology Voronoi treemap and the ontology replace functionality combine the keywords with ontological knowledge the model cannot deliver.

\paragraph{Providing Augmentations through Modular Widgets}
Different tools and augmentations are relevant depending on the tasks a user wants to solve.
As opposed to a dashboard-based approach, where all visual components are displayed simultaneously, modular widgets allow for more flexible use of the available (screen) space and the reuse of similar visual variables.
This, in return, requires careful categorization of the available widgets and useful presets for each task so that visual variables (e.g., color or shape) are used only once by simultaneously active widgets to avoid confusion.

\paragraph{Usefulness for Non-Technical Users and Linguistic Experts}
As our evaluation shows, the aforementioned mechanisms enable powerful modes of LLM output analysis.
Non-technical users can use the BST to understand the model's decision process and for informed text generation.
Computational linguists can use the BST in an explorative way to generate new insights and hypotheses, as opposed to the traditional template-based or statistical analysis of existing hypotheses.

\subsection{Limitations and Future Work}

\paragraph{Applicability to State-of-the-Art Models}
In this work, we demonstrate our approach using GPT2 and Bloom.
Beyond that, \citet{Spinner2023RevealingUnwrittenVisual} show how generAItor can be used to generate meaningful linguistic insights for different models, including GPT2, Bloom, RedPajama Base, and RedPajama Instruct~\cite{Computer2023RedpajamaOpenSource}.
We observe that our approach becomes more potent with larger models as the output diversity increases and the alternatives in the BST become more meaningful.
In general, our approach applies to causal language transformers if they
(1) provide access to the high-dimensional token-wise embeddings and
(2) output the probabilities of the next top-$k$ tokens.
While the second requirement is imperative to generate the BST, the first requirement is only needed for the embedding-based widgets.

This means that large parts of our approach are transferable to GPT4 as the current state-of-the-art in causal language modeling.
The \href{https://platform.openai.com/docs/api-reference/completions/create\#completions/create-logprobs}{OpenAI API} provides access to the logprobs of the top-$k$ tokens, which can be used to generate the BST.
Despite the high-dimensional embeddings not being available for GPT4, the embedding widgets can still be powered from the embeddings produced by other transformers.
\citet{Sevastjanova2022LmfingerprintsVisualExplanations} and \citet{Kehlbeck2021DemystifyingEmbeddingSpace} have studied the embedding spaces of prominent transformers, suggesting that using the token embeddings of other models might even be beneficial for semantic token analysis.

\paragraph{Transfer of Our Proposed Techniques to Existing Interfaces}
Our approach targets specific user groups.
However, we envision some means of explainability embedded into the prominent chat-- and completion-based interfaces, like ChatGPT or GitHub Copilot\footnote{\url{https://github.com/features/copilot}}.
Currently, ChatGPT only outputs text, and each adaptation has to be triggered by refining the prompt in the hope that the desired output will be generated.
This can be frustrating, especially for hallucinated text parts, where no easy solution for editing is available.
Here, showing alternative outputs and providing the user with explainability on the likeliness of sequences could bring huge advantages.
While GitHub Copilot does show alternatives, those alternatives remain unexplained.
Here, showing probabilities or annotating structural elements, c.f., keyword extraction (\cref{subsec:bst}) and --coloring~\refwidget{EM}{\inlinegraphicsm{1em}{-0.1em}{images/icons/widget-colored/2d-embedding-map}}, could further improve the usefulness.

\paragraph{Bridging Between Explorative and Statistical Analysis}
Our approach is explorative in nature, allowing users to generate new hypotheses and insights.
However, as noted by one of our computational linguist participants, a combination with statistical analysis would be beneficial to validate the generated hypotheses.
Therefore, we envision a tighter integration of our approach with statistical analysis tools, e.g., to validate the generated hypotheses with statistical tests.
Once this integration is established, annotating the BST branches with statistical metrics could bridge the gap between explorative and statistical analysis.
For the current version of the system, we decided against annotating the branches with linguistic metrics to prevent the user from drawing false generalizations from local observations.

\paragraph{Support for Model Developers}
Our interface also provides information relevant to model developers.
However, for model debugging and refinement, additional tools, e.g., to observe the effects of fine-tuning or investigate common errors in model and data, might be needed.

\paragraph{Extension to Other Tasks and User Groups}
The presented widgets are well-rounded for the described tasks and target user groups.
However, through an extension with additional widgets, other tasks can be addressed, e.g., informed text summarization for students.

\paragraph{Comparison Across Models}
While our approach allows loading different generative language transformers, comparative analysis is yet only possible between prompts.
However, this is not a limitation of our proposed tree-in-the-loop approach and will be implemented in future iterations of the system, enabling additional modes of analysis.

%% file: sections/conclusion.tex
We present the tree-in-the-loop paradigm, putting the beam search tree in the center of the generAItor Visual Analytics technique for language model explainability, comparability, and adaptability.
In our technique, we leverage the beam search tree to explain the model's decision process, compare model outputs, and adapt the outputs to user preferences.
Enhancing the tree with task-specific widgets creates synergies between the tree and targeted visualizations, interactions, and in-situ explanations.
Finally, we provide a three-fold evaluation of our approach.
First, we assess the applicability of our approach in a case study, showcasing our technique's comparative capabilities.
Particularly, we show how the interplay between the beam search tree and widgets enables new analysis modes, leading to interesting linguistic insights on model biases.
Second, we perform two qualitative user studies, the first with six non-experts and the second with four computational linguists, proving the usability of our approach for text generation tasks and linguistic analyses.
Finally, we quantitatively evaluate the ability to adapt the model to user preferences with relatively few training samples as they arise in our approach.

%% file: appendix/pipelines.tex
This section explains the pipelines that have been implemented to provide the functionalities of generAItor.

\subsection{Natural Language Generation Pipeline}
\label{apx:subsec:pipelines_nlg}
We generate text by using the beam search algorithm, always following the prediction with the highest probability.
The resulting beam search tree is stored as a graph in the backend of our application.
All functionalities of our system use, augment, or modify the tree.
In the following, we describe the different pipelines updating the tree state.

\paragraph{Prediction Pipeline}
We use the tokenized beam sequence from the root node up to the \HEAD node as the model input for the prediction, truncated to GPT-2's maximal sequence length of $l_{\max} = 1024$.
Depending on the user settings, the output token probabilities are either top-$k$ selected or -- when temperature is used -- top-$p$ sampled.
Finally, we append the new tokens to the beam search tree.
The full \textcolor{PredictionPipelineColor}{\textbf{Prediction Pipeline}} is depicted in~\cref{fig:nlp-pipeline}.

\paragraph{Keyword Extraction \& --Coloring}
We use YAKE~\cite{Campos2020YakeKeywordExtraction} to automatically extract keywords of an $n$-gram size of $n=1$ from the beam search tree's sequences.
Next, we tokenize the extracted keywords using the GPT-2 tokenizer, pass them to the GPT-2 model and extract the high-dimensional embeddings from GPT-2's layer $11$, maximizing the surrounding context captured by the embeddings~\cite{Sevastjanova2022LmfingerprintsVisualExplanations}.
Note that the keywords extracted by YAKE often consist of multiple split-tokens, e.g., when the keyword is a proper noun.
In this case, we average the high-dimensional embeddings of the split tokens.
To reduce the dimensionality of the embeddings from $768$ to $2$, we use a UMAP~\cite{McInnes2018UmapUniformManifold} projection pre-fitted onto keywords extracted from the MultiNLI dataset~\cite{Williams2018BroadCoverageChallenge}.
The now two-dimensional projected embedding vectors are normalized and used to sample a color on a two-dimensional colormap~\cite{Steiger2015ExplorativeAnalysis2d}.
The full \textcolor{KwEmbeddingPipelineColor}{\textbf{Keyword Embedding Pipeline}} is shown in \cref{fig:nlp-pipeline}.

\subsection{BabelNet Embedding Pipeline}
\label{apx:subsec:pipelines_babelnet}
To build the ontology graph, we leverage the power of a semantic network (BabelNet~\cite{Navigli2012BabelnetAutomaticConstruction}) and its adjacent disambiguation API (Babelfy~\cite{Moro2014EntityLinkingMeets}).
First, each keyword from the beam search tree is disambiguated in context using the Babelfy API.
The resulting BabelNet Synset is used to query a BabelNet Index v5.1.
To create a unified ontology graph, part-of-speech (POS) tags have to be considered, as the hypernym hierarchies inside BabelNet are disconnected for each POS tag.
Therefore, we must expand each keyword with a set of potential synset nouns that represent it best.
We then build and grow the ontology graph, starting with the keywords as leaf nodes.
The keywords are attached to their expanded synsets and we traverse their hypernym relations upwards.
The higher in the hierarchy a synset is, the more abstract it will be.
Therefore, at some point, the synsets are not conveying helpful information to the user.
Instead, it would make sense to reduce the hypernym relation at some point.
This decision is made using another attribute that exists on many BabelNet synsets---its BabelDomain~\cite{CamachoCollados2017BabeldomainsLargeScale}.
Domains are general groups of words that share a similarity or concept.
They are available for many synsets.
The domains of BabelNet often cover several concepts, such as Biology.
We split each domain into a collection of subdomains (BIOLOGY - Animal, Person).
If a synset does not have a domain, we stop traversing the hypernym relations and instead attach the synset to its most similar subdomain and domain.
The ontology graph can grow large quickly, as the hypernym relations are often intertwined and contain many synsets.
To simplify the tree, we remove nodes that only act as connecting nodes between two synsets.
The result is a relatively compact collection of trees, with one tree for each domain.
When predictions are made, the initial ontology graph is expanded with new keywords.
Visualizing this ontology graph directly can create large trees, as multiple instances of the same keyword appear multiple times, creating a multitude of leaf nodes.
We therefore instead simplify the graph further into four distinct layers, where each node can only have one parent relation.
This graph can then be visualized using a Voronoi diagram.
We use the D3 Voronoi treemap ~\footnote{\url{https://github.com/Kcnarf/d3-voronoi-treemap}} implementation to create a Voronoi treemap of the hierarchy and allow the user to select the layer they want to view.
As the upper layers aggregate the keywords to the same synset, they offer a more compact view of the domains and keywords of the prediction graph.
The \textcolor{BabelnetEmbeddingPipelineColor}{\textbf{BabelNet Embedding Pipeline}} is shown in \cref{fig:nlp-pipeline-bert-predictions}.

\subsection{Masked Ontological Replacement Pipeline}
\label{apx:subsec:ontological-replace}
To create the domain-specific, context-sensitive suggestions of the ontology replace function, we combine the power of the semantic network with masked language modeling.
The goal is to replace a specific word with another suggestion that fits its context and can be grouped into domains.
To solve this, we use a combination of BERT and ARES Embeddings~\cite{Scarlini2020MoreContextsComes}.
ARES embeddings are powerful sense embeddings with high-dimensional representatives for all WordNet synsets.
They were trained in a semi-supervised approach combining a lexical knowledge base with BERT Large embeddings
and place WordNet synsets in the same embedding space as BERT embeddings.
This way, for a given WordNet synset, we can query the closest BERT embedding and vice versa.
Because BabelNet has WordNet bindings for many BabelNet synsets, we assign each subdomain a BabelNet and their respective WordNet synset.
This way, each subdomain can be assigned to an embedding vector via ARES.
The \textcolor{OntoReplacePipelineColor}{\textbf{Masked Ontological Replacement Pipeline}} can be observed in \cref{fig:nlp-pipeline-onto-tree}.
For each keyword in the Beam Search Tree, we take the word and its sentence and replace it with the $[MASK]$ token.
Afterwards, we can use top-$k$ prediction on BERT to query a large number of predictions that would otherwise be impossible to show the user in a compact way ($k=200$).
We tokenize each predicted word and extract the model logits in context, extracting and squeezing layers 8-11, which are then appended to match the ARES embeddings length~($n=2048$).
After this step, we have a set of embeddings for subdomains in the ontology graph and a set of embeddings for the predictions in the beam search tree.
To bring them together, we look for the nearest neighbors of all embedding vectors.
To speed up the process, we created a custom FAISS~\cite{Johnson2019BillionScaleSimilarity} index, which we can use to query nearest neighbors efficiently.
Subdomains and predictions are matched via their overlapping nearest neighbors.
The resulting predictions are then attached to each keyword and shown on demand via the ontology replace function.

\begin{landscape}
    \begin{figure}
        \centering
        \includegraphics[width=\linewidth]{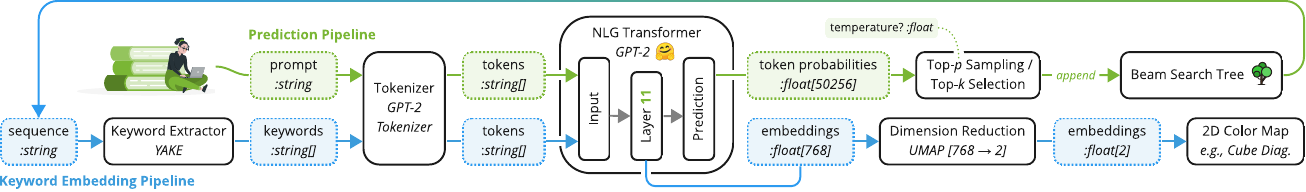}
        \caption{The pipeline to expand the beam search tree and assign the semantic keyword color information to its nodes.}
        \label{fig:nlp-pipeline}
    \end{figure}

    \begin{figure}
        \centering
        \includegraphics[width=0.6\linewidth]{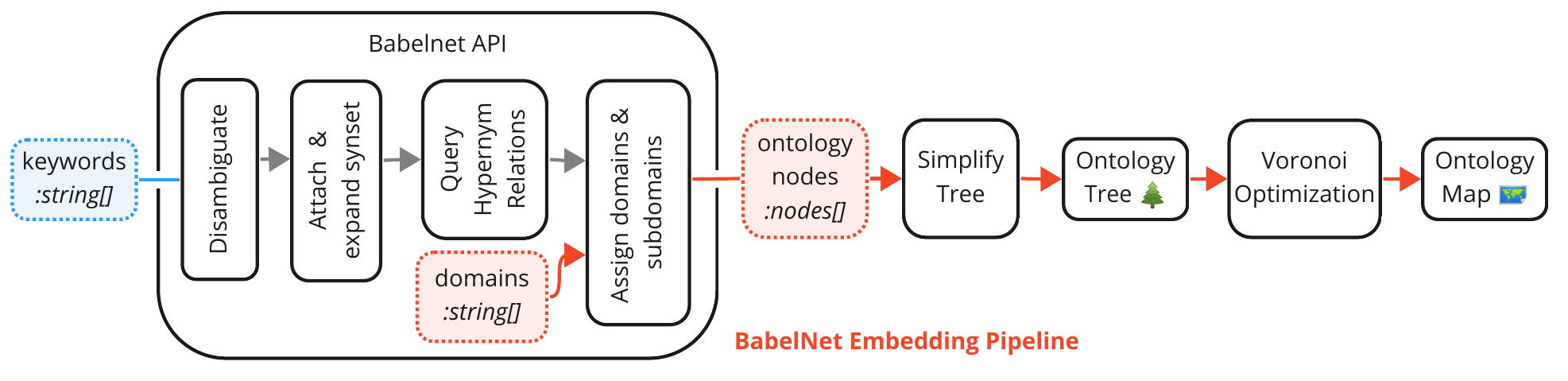}
        \caption{Keywords are attached to the ontology graph via the BabelNet embedding pipeline. This graph is then further simplified and the hierarchy is used to create an Ontology Map using a Voronoi diagram visualization.}
        \label{fig:nlp-pipeline-bert-predictions}
    \end{figure}

    \begin{figure}
        \centering
        \includegraphics[width=\linewidth]{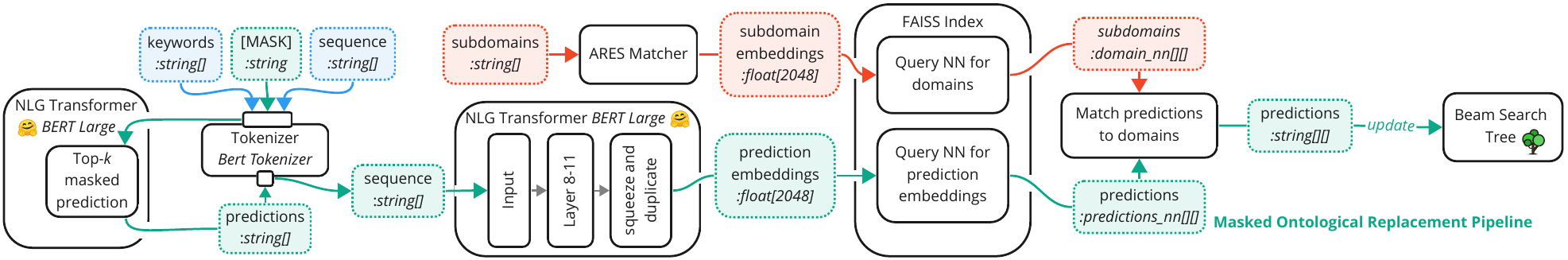}
        \caption{Domain-specific keywords are attached to each node of the beam search tree by comparing the nearest neighbours of the domain's ARES embeddings and the nearest neighbours of the BERT predictions that could replace the keyword of the node.}
        \label{fig:nlp-pipeline-onto-tree}
    \end{figure}
\end{landscape}

%% file: literature/references.bib
@Article{Alba2022OpenaiChatbotSpits,
  author    = {Alba, Davey},
  journal   = {Bloomberg},
  title     = {{OpenAI Chatbot Spits Out Biased Musings, Despite Guardrails}},
  year      = {2022},
  publisher = {Bloomberg},
  url       = {https://www.bloomberg.com/news/newsletters/2022-12-08/chatgpt-open-ai-s-chatbot-is-spitting-out-biased-sexist-results},
  urldate   = {2023-03-30},
}

@InProceedings{Alnegheimish2022UsingNaturalSentence,
  author    = {Alnegheimish, Sarah and Guo, Alicia and Sun, Yi},
  booktitle = {Proceedings of the Conference of the North American Chapter of the Association for Computational Linguistics: Human Language Technologies},
  title     = {Using Natural Sentence Prompts for Understanding Biases in Language Models},
  year      = {2022},
  address   = {Seattle, United States},
  pages     = {2824--2830},
  publisher = {Association for Computational Linguistics},
}

@InProceedings{AriasHernandez2011PairAnalyticsCapturing,
  author    = {R Arias-Hernandez and L T Kaastra and T M Green and B Fisher},
  booktitle = {Hawaii International Conference on System Sciences},
  title     = {Pair Analytics: Capturing Reasoning Processes in Collaborative Visual Analytics},
  year      = {2011},
  publisher = {{IEEE}},
}

@Article{Bahdanau2014NeuralMachineTranslation,
  author        = {Dzmitry Bahdanau and Kyunghyun Cho and Yoshua Bengio},
  title         = {Neural Machine Translation by Jointly Learning to Align and Translate},
  year          = {2014},
  archiveprefix = {arXiv},
  eprint        = {1409.0473},
}

@Article{Bengio2000NeuralProbabilisticLanguage,
  author  = {Bengio, Yoshua and Ducharme, R{\'e}jean and Vincent, Pascal},
  journal = {Advances in Neural Information Processing Systems},
  title   = {A neural probabilistic language model},
  year    = {2000},
  volume  = {13},
}

@InProceedings{Blodgett2020LanguageTechnologyIs,
  author    = {Blodgett, Su Lin and Barocas, Solon and Daum{\'e} III, Hal and Wallach, Hanna},
  booktitle = {Proceedings of the Association for Computational Linguistics},
  title     = {Language (Technology) is Power: A Critical Survey of {``}Bias{''} in {NLP}},
  year      = {2020},
  address   = {Online},
  pages     = {5454--5476},
  publisher = {Association for Computational Linguistics},
}

@Article{Brown2020LanguageModelsAre,
  author        = {Tom B. Brown and Benjamin Mann and Nick Ryder and Melanie Subbiah and Jared Kaplan and Prafulla Dhariwal and Arvind Neelakantan and Pranav Shyam and Girish Sastry and Amanda Askell and Sandhini Agarwal and Ariel Herbert-Voss and Gretchen Krueger and Tom Henighan and Rewon Child and Aditya Ramesh and Daniel M. Ziegler and Jeffrey Wu and Clemens Winter and Christopher Hesse and Mark Chen and Eric Sigler and Mateusz Litwin and Scott Gray and Benjamin Chess and Jack Clark and Christopher Berner and Sam McCandlish and Alec Radford and Ilya Sutskever and Dario Amodei},
  title         = {Language Models are Few-Shot Learners},
  year          = {2020},
  archiveprefix = {arXiv},
  eprint        = {2005.14165},
}

@Article{Caliskan2017SemanticsDerivedAutomatically,
  author  = {Aylin Caliskan and Joanna J. Bryson and Arvind Narayanan},
  journal = {Science},
  title   = {Semantics derived automatically from language corpora contain human-like biases},
  year    = {2017},
  number  = {6334},
  pages   = {183-186},
  volume  = {356},
}

@InProceedings{CamachoCollados2017BabeldomainsLargeScale,
  author    = {Jose Camacho-Collados and Roberto Navigli},
  booktitle = {Proceedings of the 15th Conference of the European Chapter of the Association for Computational Linguistics: Short Papers},
  title     = {{BabelDomains}: Large-Scale Domain Labeling of Lexical Resources},
  year      = {2017},
  publisher = {Association for Computational Linguistics},
  volume    = {2},
}

@Article{Campos2020YakeKeywordExtraction,
  author    = {Ricardo Campos and Vítor Mangaravite and Arian Pasquali and Alípio Jorge and Célia Nunes and Adam Jatowt},
  journal   = {Information Sciences},
  title     = {{YAKE}! Keyword extraction from single documents using multiple local features},
  year      = {2020},
  pages     = {257--289},
  volume    = {509},
  publisher = {Elsevier {BV}},
}

@InProceedings{Chen2021ConstructingTaxonomiesPretrained,
  author    = {Catherine Chen and Kevin Lin and Dan Klein},
  booktitle = {Proceedings of the Conference of the North American Chapter of the Association for Computational Linguistics: Human Language Technologies},
  title     = {Constructing Taxonomies from Pretrained Language Models},
  year      = {2021},
  publisher = {Association for Computational Linguistics},
}

@Misc{Computer2023RedpajamaOpenSource,
  author       = {Together Computer},
  howpublished = {\url{https://github.com/togethercomputer/RedPajama-Data}},
  title        = {RedPajama: An Open Source Recipe to Reproduce LLaMA training dataset},
  year         = {2023},
}

@InProceedings{Conia2020ConceptionMultilinguallyEnhanced,
  author    = {Simone Conia and Roberto Navigli},
  booktitle = {Proceedings of the 28th International Conference on Computational Linguistics},
  title     = {Conception: Multilingually-Enhanced, Human-Readable Concept Vector Representations},
  year      = {2020},
  publisher = {International Committee on Computational Linguistics},
}

@Book{Corver2001SemiLexicalCategories,
  author    = {Norbert Corver and Henk van Riemsdijk},
  publisher = {De Gruyter Mouton},
  title     = {Semi-lexical Categories: The Function of Content Words and the Content of Function Words},
  year      = {2001},
  address   = {Berlin, New York},
}

@InProceedings{Danilevsky2020SurveyStateExplainable,
  author    = {Danilevsky, Marina and Qian, Kun and Aharonov, Ranit and Katsis, Yannis and Kawas, Ban and Sen, Prithviraj},
  booktitle = {Proceedings of the 1st Conference of the Asia-Pacific Chapter of the Association for Computational Linguistics and the 10th International Joint Conference on Natural Language Processing},
  title     = {A Survey of the State of Explainable {AI} for Natural Language Processing},
  year      = {2020},
  address   = {Suzhou, China},
  pages     = {447--459},
  publisher = {Association for Computational Linguistics},
}

@Article{Dathathri2019PlugPlayLanguage,
  author    = {Dathathri, Sumanth and Madotto, Andrea and Lan, Janice and Hung, Jane and Frank, Eric and Molino, Piero and Yosinski, Jason and Liu, Rosanne},
  title     = {Plug and Play Language Models: A Simple Approach to Controlled Text Generation},
  year      = {2019},
  copyright = {arXiv.org perpetual, non-exclusive license},
  publisher = {arXiv},
  url       = {https://arxiv.org/abs/1912.02164},
}

@Misc{DeepNLP2023BiasNlp,
  author  = {{Deep NLP}},
  note    = {[Online; accessed 15. Nov. 2023]},
  title   = {Bias in {NLP}},
  year    = {2023},
  journal = {GitHub},
  url     = {https://github.com/cisnlp/bias-in-nlp},
  urldate = {2023-11-15},
}

@Article{Devlin2018BertPreTraining,
  author        = {Jacob Devlin and Ming-Wei Chang and Kenton Lee and Kristina Toutanova},
  title         = {BERT: Pre-training of Deep Bidirectional Transformers for Language Understanding},
  year          = {2018},
  archiveprefix = {arXiv},
  eprint        = {1810.04805},
}

@InProceedings{Du2022ReadReviseRepeat,
  author    = {Wanyu Du and Zae Myung Kim and Vipul Raheja and Dhruv Kumar and Dongyeop Kang},
  booktitle = {Proceedings of the First Workshop on Intelligent and Interactive Writing Assistants},
  title     = {Read, Revise, Repeat: A System Demonstration for Human-in-the-loop Iterative Text Revision},
  year      = {2022},
  publisher = {Association for Computational Linguistics},
}

@Article{ElAssady2018ThreadreconstructorModelingReply,
  author    = {Mennatallah El-Assady and Rita Sevastjanova and Daniel Keim and Christopher Collins},
  journal   = {Computer Graphics Forum},
  title     = {{ThreadReconstructor}: Modeling Reply-Chains to Untangle Conversational Text through Visual Analytics},
  year      = {2018},
  number    = {3},
  pages     = {351--365},
  volume    = {37},
  publisher = {Wiley},
}

@InProceedings{ElAssady2019TowardsXaiStructuring,
  author    = {El-Assady, M. and Jentner, W. and Kehlbeck, R. and Schlegel, U. and Sevastjanova, R. and Sperrle, F. and Spinner, T. and Keim, D.},
  booktitle = {{ACM CHI 2019 Workshop: Human--Centered Machine Learning Perspectives}},
  title     = {{Towards XAI: Structuring the Processes of Explanations}},
  year      = {2019},
  ids       = {strategies2019},
}

@Article{ElAssady2022SemanticColorMapping,
  author  = {Mennatallah El-Assady and Rebecca Kehlbeck and Yannick Metz and Udo Schlegel and Rita Sevastjanova and Fabian Sperrle and Thilo Spinner},
  journal = {4th IEEE Workshop on Visualization Guidelines in Research, Design, and Education},
  title   = {Semantic Color Mapping: A Pipeline for Assigning Meaningful Colors to Text},
  year    = {2022},
}

@InProceedings{Ethayarajh2019HowContextualAre,
  author    = {Ethayarajh, Kawin},
  booktitle = {Proceedings of the Conference on Empirical Methods in Natural Language Proceedings and the International Joint Conference on Natural Language Processing},
  title     = {{How Contextual are Contextualized Word Representations? {C}omparing the Geometry of {BERT}, {ELM}o, and {GPT}-2 Embeddings}},
  year      = {2019},
  address   = {Hong Kong, China},
  pages     = {55--65},
  publisher = {ACL},
}

@Article{GarridoMunoz2021SurveyBiasDeep,
  author    = {Garrido-Mu{\~n}oz, Ismael and Montejo-R{\'a}ez, Arturo and Mart{\'i}nez-Santiago, Fernando and Ure{\~n}a-L{\'o}pez, L Alfonso},
  journal   = {Applied Sciences},
  title     = {A survey on bias in deep NLP},
  year      = {2021},
  number    = {7},
  pages     = {3184},
  volume    = {11},
  publisher = {Multidisciplinary Digital Publishing Institute},
}

@Article{Gatt2018SurveyStateArt,
  author    = {Albert Gatt and Emiel Krahmer},
  journal   = {Journal of Artificial Intelligence Research},
  title     = {Survey of the State of the Art in Natural Language Generation: Core tasks, applications and evaluation},
  year      = {2018},
  pages     = {65--170},
  volume    = {61},
  publisher = {{AI} Access Foundation},
}

@Article{Gehrmann2019VisualInteractionDeep,
  author    = {Sebastian Gehrmann and Hendrik Strobelt and Robert Kruger and Hanspeter Pfister and Alexander M. Rush},
  journal   = {{IEEE} Transactions on Visualization and Computer Graphics},
  title     = {Visual Interaction with Deep Learning Models through Collaborative Semantic Inference},
  year      = {2019},
  pages     = {1--1},
  publisher = {Institute of Electrical and Electronics Engineers ({IEEE})},
}

@Article{Hartmann2021PowerBrandSelfies,
  author  = {Hartmann, Jochen and Heitmann, Mark and Schamp, Christina and Netzer, Oded},
  journal = {Journal of Marketing Research},
  title   = {The Power of Brand Selfies},
  year    = {2021},
}

@InProceedings{He2021ParallelRefinementsLexically,
  author    = {Xingwei He},
  booktitle = {Proceedings of the Conference on Empirical Methods in Natural Language Processing},
  title     = {Parallel Refinements for Lexically Constrained Text Generation with {BART}},
  year      = {2021},
  publisher = {Association for Computational Linguistics},
}

@InProceedings{Howard2018UniversalLanguageModel,
  author    = {Howard, Jeremy and Ruder, Sebastian},
  booktitle = {Proceedings of the 56th Annual Meeting of the Association for Computational Linguistics},
  title     = {Universal Language Model Fine-tuning for Text Classification},
  year      = {2018},
  address   = {Melbourne, Australia},
  pages     = {328--339},
  publisher = {Association for Computational Linguistics},
  volume    = {1},
}

@InProceedings{Hu2017ControlledGenerationText,
  author    = {Zhiting Hu and Zichao Yang and Xiaodan Liang and Ruslan Salakhutdinov and Eric P. Xing},
  booktitle = {Proceedings of the 34th International Conference on Machine Learning},
  title     = {Toward Controlled Generation of Text},
  year      = {2017},
  editor    = {Precup, Doina and Teh, Yee Whye},
  pages     = {1587--1596},
  publisher = {PMLR},
  series    = {Proceedings of Machine Learning Research},
  volume    = {70},
}

@InProceedings{Hua2020PairPlanningIterative,
  author    = {Xinyu Hua and Lu Wang},
  booktitle = {Proceedings of the Conference on Empirical Methods in Natural Language Processing ({EMNLP})},
  title     = {{PAIR}: Planning and Iterative Refinement in Pre-trained Transformers for Long Text Generation},
  year      = {2020},
  publisher = {Association for Computational Linguistics},
}

@InProceedings{Huang2020CorelSeedGuided,
  author    = {Jiaxin Huang and Yiqing Xie and Yu Meng and Yunyi Zhang and Jiawei Han},
  booktitle = {Proceedings of the 26th {ACM} {SIGKDD} International Conference on Knowledge Discovery and Data Mining},
  title     = {{CoRel}: Seed-Guided Topical Taxonomy Construction by Concept Learning and Relation Transferring},
  year      = {2020},
  publisher = {{ACM}},
}

@Article{Jelinek1977PerplexityAMeasureDifficulty,
  author    = {Jelinek, Fred and Mercer, Robert L and Bahl, Lalit R and Baker, James K},
  journal   = {The Journal of the Acoustical Society of America},
  title     = {Perplexity—a measure of the difficulty of speech recognition tasks},
  year      = {1977},
  number    = {S1},
  pages     = {S63--S63},
  volume    = {62},
  publisher = {Acoustical Society of America},
}

@Article{Ji2023SurveyHallucinationNatural,
  author    = {Ziwei Ji and Nayeon Lee and Rita Frieske and Tiezheng Yu and Dan Su and Yan Xu and Etsuko Ishii and Ye Jin Bang and Andrea Madotto and Pascale Fung},
  journal   = {{ACM} Computing Surveys},
  title     = {Survey of Hallucination in Natural Language Generation},
  year      = {2023},
  number    = {12},
  pages     = {1--38},
  volume    = {55},
  publisher = {Association for Computing Machinery ({ACM})},
}

@InProceedings{Jiang2022TaxoenrichSelfSupervised,
  author    = {Minhao Jiang and Xiangchen Song and Jieyu Zhang and Jiawei Han},
  booktitle = {Proceedings of the {ACM} Web Conference},
  title     = {{TaxoEnrich}: Self-Supervised Taxonomy Completion via Structure-Semantic Representations},
  year      = {2022},
  publisher = {{ACM}},
}

@Article{Johnson2019BillionScaleSimilarity,
  author    = {Johnson, Jeff and Douze, Matthijs and J{\'e}gou, Herv{\'e}},
  journal   = {IEEE Transactions on Big Data},
  title     = {Billion-scale similarity search with {GPUs}},
  year      = {2019},
  number    = {3},
  pages     = {535--547},
  volume    = {7},
  publisher = {IEEE},
}

@InProceedings{Kalouli2022NegationCoordinationQuantifiers,
  author    = {Kalouli, Aikaterini-Lida and Sevastjanova, Rita and Beck, Christin and Romero, Maribel},
  booktitle = {Proceedings of the 29th International Conference on Comp. Ling.},
  title     = {Negation, Coordination, and Quantifiers in Contextualized Language Models},
  year      = {2022},
  address   = {Gyeongju, Republic of Korea},
  pages     = {3074--3085},
  publisher = {International Committee on Computational Linguistics},
}

@Article{Kehlbeck2021DemystifyingEmbeddingSpace,
  author  = {Rebecca Kehlbeck and Rita Sevastjanova and Thilo Spinner and Tobias Stähle and Mennatallah El-Assady},
  journal = {Proceedings of the Workshop on Visualization for AI Explainability (VISxAI)},
  title   = {Demystifying the Embedding Space of Language Models},
  year    = {2021},
  note    = {\url{https://bert-vs-gpt2.dbvis.de/}},
}

@InProceedings{Lauscher2021SustainableModularDebiasing,
  author    = {Lauscher, Anne and Lueken, Tobias and Glava{\v{s}}, Goran},
  booktitle = {Findings of the Association for Computational Linguistics: EMNLP},
  title     = {Sustainable Modular Debiasing of Language Models},
  year      = {2021},
  address   = {Punta Cana, Dominican Republic},
  pages     = {4782--4797},
  publisher = {Association for Computational Linguistics},
}

@Unpublished{LeCun2023DoLanguageModels,
  author = {Yann LeCun},
  note   = {The Philosophy of Deep Learning},
  title  = {Do Language Models Need Sensory Grounding for Meaning and Understanding?},
  year   = {2023},
  url    = {https://drive.google.com/file/d/1BU5bV3X5w65DwSMapKcsr0ZvrMRU_Nbi},
}

@InProceedings{Lee2017InteractiveVisualizationManipulation,
  author    = {Lee, Jaesong and Shin, Joong-Hwi and Kim, Jun-Seok},
  booktitle = {Proceedings of the Conference on Empirical Methods in Natural Language Processing: System Demonstrations},
  title     = {Interactive Visualization and Manipulation of Attention-based Neural Machine Translation},
  year      = {2017},
  address   = {Copenhagen, Denmark},
  pages     = {121--126},
  publisher = {Association for Computational Linguistics},
}

@InProceedings{Lewis2020BartDenoisingSequence,
  author    = {Lewis, Mike and Liu, Yinhan and Goyal, Naman and Ghazvininejad, Marjan and Mohamed, Abdelrahman and Levy, Omer and Stoyanov, Veselin and Zettlemoyer, Luke},
  booktitle = {Proceedings of the 58th Annual Meeting of the Association for Computational Linguistics},
  title     = {{BART}: Denoising Sequence-to-Sequence Pre-training for Natural Language Generation, Translation, and Comprehension},
  year      = {2020},
  address   = {Online},
  pages     = {7871--7880},
  publisher = {Association for Computational Linguistics},
}

@Article{Lex2014UpsetVisualizationIntersecting,
  author    = {Alexander Lex and Nils Gehlenborg and Hendrik Strobelt and Romain Vuillemot and Hanspeter Pfister},
  journal   = {{IEEE} Transactions on Visualization and Computer Graphics},
  title     = {{UpSet}: Visualization of Intersecting Sets},
  year      = {2014},
  number    = {12},
  pages     = {1983--1992},
  volume    = {20},
  publisher = {Institute of Electrical and Electronics Engineers ({IEEE})},
}

@InProceedings{Li2021PretrainedLanguageModel,
  author    = {Junyi Li and Tianyi Tang and Wayne Xin Zhao and Ji-Rong Wen},
  booktitle = {Proceedings of the 30th International Joint Conference on Artificial Intelligence},
  title     = {Pretrained Language Model for Text Generation: A Survey},
  year      = {2021},
  publisher = {International Joint Conference on Artificial Intelligence Organization},
}

@InProceedings{Li2022Taxotrans,
  author    = {Zhuliu Li and Yiming Wang and Xiao Yan and Weizhi Meng and Yanen Li and Jaewon Yang},
  booktitle = {Proceedings of the 28th {ACM} {SIGKDD} Conference on Knowledge Discovery and Data Mining},
  title     = {{TaxoTrans}},
  year      = {2022},
  publisher = {{ACM}},
}

@InProceedings{Liang2021TowardsUnderstandingMitigating,
  author       = {Liang, Paul Pu and Wu, Chiyu and Morency, Louis-Philippe and Salakhutdinov, Ruslan},
  booktitle    = {International Conference on Machine Learning},
  title        = {Towards understanding and mitigating social biases in language models},
  year         = {2021},
  organization = {PMLR},
  pages        = {6565--6576},
}

@Article{Loshchilov2017FixingWeightDecay,
  author        = {Ilya Loshchilov and Frank Hutter},
  journal       = {CoRR},
  title         = {Fixing Weight Decay Regularization in Adam},
  year          = {2017},
  volume        = {abs/1711.05101},
  archiveprefix = {arXiv},
  bibsource     = {dblp computer science bibliography, https://dblp.org},
  eprint        = {1711.05101},
  url           = {http://arxiv.org/abs/1711.05101},
}

@Article{Lu2020GenderBiasNeural,
  author    = {Lu, Kaiji and Mardziel, Piotr and Wu, Fangjing and Amancharla, Preetam and Datta, Anupam},
  journal   = {Logic, Language, and Security: Essays Dedicated to Andre Scedrov on the Occasion of His 65th Birthday},
  title     = {Gender bias in neural natural language processing},
  year      = {2020},
  pages     = {189--202},
  publisher = {Springer},
}

@InProceedings{Maas2011LearningWordVectors,
  author    = {Maas, Andrew L. and Daly, Raymond E. and Pham, Peter T. and Huang, Dan and Ng, Andrew Y. and Potts, Christopher},
  booktitle = {Proceedings of the 49th Annual Meeting of the Association for Computational Linguistics: Human Language Technologies},
  title     = {Learning Word Vectors for Sentiment Analysis},
  year      = {2011},
  address   = {Portland, Oregon, USA},
  pages     = {142--150},
  publisher = {Association for Computational Linguistics},
}

@Article{McInnes2018UmapUniformManifold,
  author  = {McInnes, Leland and Healy, John and Saul, Nathaniel and Grossberger, Lukas},
  journal = {The Journal of Open Source Software},
  title   = {UMAP: Uniform Manifold Approximation and Projection},
  year    = {2018},
  number  = {29},
  pages   = {861},
  volume  = {3},
}

@Article{Mehrabi2021SurveyBiasFairness,
  author    = {Mehrabi, Ninareh and Morstatter, Fred and Saxena, Nripsuta and Lerman, Kristina and Galstyan, Aram},
  journal   = {ACM Computing Surveys},
  title     = {A survey on bias and fairness in machine learning},
  year      = {2021},
  number    = {6},
  pages     = {1--35},
  volume    = {54},
  publisher = {ACM New York, NY, USA},
}

@Article{Metz2022NewChatbotsCould,
  author    = {Metz, Cade},
  journal   = {New York Times},
  title     = {{The New Chatbots Could Change the World. Can You Trust Them?}},
  year      = {2022},
  issn      = {0362-4331},
  publisher = {The New York Times},
  url       = {https://www.nytimes.com/2022/12/10/technology/ai-chat-bot-chatgpt.html},
}

@InProceedings{Mishra2022CrossTaskGeneralization,
  author    = {Swaroop Mishra and Daniel Khashabi and Chitta Baral and Hannaneh Hajishirzi},
  booktitle = {Proceedings of the 60th Annual Meeting of the Association for Computational Linguistics},
  title     = {Cross-Task Generalization via Natural Language Crowdsourcing Instructions},
  year      = {2022},
  publisher = {Association for Computational Linguistics},
}

@Article{Moro2014EntityLinkingMeets,
  author    = {Andrea Moro and Alessandro Raganato and Roberto Navigli},
  journal   = {Transactions of the Association for Computational Linguistics},
  title     = {Entity Linking meets Word Sense Disambiguation: a Unified Approach},
  year      = {2014},
  pages     = {231--244},
  volume    = {2},
  publisher = {{MIT} Press - Journals},
}

@InProceedings{Nadeem2021StereosetMeasuringStereotypical,
  author    = {Nadeem, Moin and Bethke, Anna and Reddy, Siva},
  booktitle = {Proceedings of the 59th Annual Meeting of the Association for Computational Linguistics and the 11th International Joint Conference on Natural Language Processing},
  title     = {{S}tereo{S}et: Measuring stereotypical bias in pretrained language models},
  year      = {2021},
  address   = {Online},
  pages     = {5356--5371},
  publisher = {Association for Computational Linguistics},
  volume    = {1},
}

@Article{Navigli2012BabelnetAutomaticConstruction,
  author    = {Roberto Navigli and Simone Paolo Ponzetto},
  journal   = {Artificial Intelligence},
  title     = {{BabelNet}: The automatic construction, evaluation and application of a wide-coverage multilingual semantic network},
  year      = {2012},
  pages     = {217--250},
  volume    = {193},
  publisher = {Elsevier {BV}},
}

@Misc{OpenAI2023Gpt4Technical,
  author        = {OpenAI},
  title         = {GPT-4 Technical Report},
  year          = {2023},
  archiveprefix = {arXiv},
  eprint        = {2303.08774},
}

@InProceedings{Ouyang2022TrainingLanguageModels,
  author    = {Ouyang, Long and Wu, Jeffrey and Jiang, Xu and Almeida, Diogo and Wainwright, Carroll and Mishkin, Pamela and Zhang, Chong and Agarwal, Sandhini and Slama, Katarina and Ray, Alex and Schulman, John and Hilton, Jacob and Kelton, Fraser and Miller, Luke and Simens, Maddie and Askell, Amanda and Welinder, Peter and Christiano, Paul F and Leike, Jan and Lowe, Ryan},
  booktitle = {Advances in Neural Information Processing Systems},
  title     = {Training language models to follow instructions with human feedback},
  year      = {2022},
  editor    = {S. Koyejo and S. Mohamed and A. Agarwal and D. Belgrave and K. Cho and A. Oh},
  pages     = {27730--27744},
  publisher = {Curran Associates, Inc.},
  volume    = {35},
}

@InProceedings{Padmakumar2022MachineLoopRewriting,
  author    = {Vishakh Padmakumar and He He},
  booktitle = {Proceedings of the Conference of the North American Chapter of the Association for Computational Linguistics: Human Language Technologies},
  title     = {Machine-in-the-Loop Rewriting for Creative Image Captioning},
  year      = {2022},
  publisher = {Association for Computational Linguistics},
}

@Article{Paulus2017DeepReinforcedModel,
  author        = {Romain Paulus and Caiming Xiong and Richard Socher},
  journal       = {CoRR},
  title         = {A Deep Reinforced Model for Abstractive Summarization},
  year          = {2017},
  volume        = {abs/1705.04304},
  archiveprefix = {arXiv},
  bibsource     = {dblp computer science bibliography, https://dblp.org},
  eprint        = {1705.04304},
  url           = {http://arxiv.org/abs/1705.04304},
}

@Misc{Platen2020HowGenerateText,
  author  = {Patrick von Platen},
  note    = {[Online; accessed 29. Mar. 2023]},
  title   = {{How to generate text: using different decoding methods for language generation with Transformers}},
  year    = {2020},
  url     = {https://huggingface.co/blog/how-to-generate},
  urldate = {2023-03-29},
}

@InProceedings{Qin2020BackFutureUnsupervised,
  author    = {Lianhui Qin and Vered Shwartz and Peter West and Chandra Bhagavatula and Jena D. Hwang and Ronan Le Bras and Antoine Bosselut and Yejin Choi},
  booktitle = {Proceedings of the Conference on Empirical Methods in Natural Language Processing ({EMNLP})},
  title     = {Back to the Future: Unsupervised Backprop-based Decoding for Counterfactual and Abductive Commonsense Reasoning},
  year      = {2020},
  publisher = {Association for Computational Linguistics},
}

@Misc{Radford2019BetterLanguageModels,
  author       = {Alec Radford and Jeffrey Wu and Dario Amodei and Daniela Amodei and Jack Clark and Miles Brundage and Ilya Sutskever},
  howpublished = {\url{https://openai.com/blog/better-language-models/}},
  note         = {{[Online; accessed 18-March-2021]}},
  title        = {Better Language Models and Their Implications},
  year         = {2019},
}

@Article{Radford2019LanguageModelsAre,
  author = {Radford, Alec and Wu, Jeff and Child, Rewon and Luan, David and Amodei, Dario and Sutskever, Ilya},
  title  = {Language Models are Unsupervised Multitask Learners},
  year   = {2019},
}

@InCollection{Reif2019VisualizingMeasuringGeometry,
  author    = {Reif, Emily and Yuan, Ann and Wattenberg, Martin and Viegas, Fernanda B and Coenen, Andy and Pearce, Adam and Kim, Been},
  booktitle = {Advances in Neural Information Processing Systems},
  publisher = {Curran Associates, Inc.},
  title     = {{Visualizing and Measuring the Geometry of {BERT}}},
  year      = {2019},
  editor    = {H. Wallach and H. Larochelle and A. Beygelzimer and F. d'Alch\'{e}-Buc and E. Fox and R. Garnett},
  pages     = {8594--8603},
}

@Article{Rogers2020PrimerBertologyWhat,
  author  = {Rogers, Anna and Kovaleva, Olga and Rumshisky, Anna},
  journal = {Transactions of the Association for Computational Linguistics},
  title   = {{A Primer in BERTology: What We Know About How BERT Works}},
  year    = {2020},
  pages   = {842-866},
  volume  = {8},
}

@Article{Roose2023HowChatbotsLarge,
  author    = {Roose, Kevin},
  journal   = {New York Times},
  title     = {{How Chatbots and Large Language Models, or LLMs, Actually Work}},
  year      = {2023},
  issn      = {0362-4331},
  publisher = {The New York Times},
  url       = {https://www.nytimes.com/2023/03/28/technology/ai-chatbots-chatgpt-bing-bard-llm.html},
  urldate   = {2023-11-03},
}

@Article{Rumelhart1986LearningRepresentationsBack,
  author    = {David E. Rumelhart and Geoffrey E. Hinton and Ronald J. Williams},
  journal   = {Cahiers De La Revue De Theologie Et De Philosophie},
  title     = {Learning representations by back-propagating errors},
  year      = {1986},
  number    = {6088},
  pages     = {533--536},
  volume    = {323},
  publisher = {Springer Science and Business Media {LLC}},
}

@InProceedings{Scarlini2020MoreContextsComes,
  author    = {Scarlini, Bianca and Pasini, Tommaso and Navigli, Roberto},
  booktitle = {Proceedings of the Conference on Empirical Methods in Natural Language Processing},
  title     = {{With More Contexts Comes Better Performance: Contextualized Sense Embeddings for All-Round Word Sense Disambiguation}},
  year      = {2020},
  publisher = {Association for Computational Linguistics},
}

@Article{Sevastjanova2022BewareRationalizationTrap,
  author  = {Rita Sevastjanova and Mennatallah El-Assady},
  journal = {Conference: Communication in Human-AI Interaction Workshop at IJCAI-ECAI},
  title   = {{Beware the Rationalization Trap! When Language Model Explainability Diverges from our Mental Models of Language}},
  year    = {2022},
  volume  = {abs/2207.06897},
}

@Article{Sevastjanova2022LmfingerprintsVisualExplanations,
  author    = {Sevastjanova, Rita and Kalouli, Aikaterini-Lida and Beck, Christin and Hauptmann, Hanna and El-Assady, Mennatallah},
  journal   = {Computer Graphics Forum},
  title     = {{LMFingerprints}: Visual Explanations of Language Model Embedding Spaces through Layerwise Contextualization Scores},
  year      = {2022},
  number    = {3},
  pages     = {295--307},
  volume    = {41},
  publisher = {Wiley},
}

@Article{Spinner2020ExplainerVisualAnalytics,
  author    = {Thilo Spinner and Udo Schlegel and Hanna Schafer and Mennatallah El-Assady},
  journal   = {{IEEE} Transactions on Visualization and Computer Graphics},
  title     = {{explAIner}: A Visual Analytics Framework for Interactive and Explainable Machine Learning},
  year      = {2020},
  number    = {1},
  volume    = {26},
  publisher = {Institute of Electrical and Electronics Engineers ({IEEE})},
}

@Misc{Spinner2023RevealingUnwrittenVisual,
  author        = {Thilo Spinner and Rebecca Kehlbeck and Rita Sevastjanova and Tobias Stähle and Daniel A. Keim and Oliver Deussen and Andreas Spitz and Mennatallah El-Assady},
  title         = {Revealing the Unwritten: Visual Investigation of Beam Search Trees to Address Language Model Prompting Challenges},
  year          = {2023},
  archiveprefix = {arXiv},
  eprint        = {2310.11252},
}

@InProceedings{Steiger2015ExplorativeAnalysis2d,
  author    = {Martin Steiger and J. Bernard and Simon Thum and Sebastian Mittelst{\"a}dt and Marco Hutter and Daniel A. Keim and J{\"o}rn Kohlhammer},
  booktitle = {WSCG},
  title     = {Explorative analysis of 2D color maps},
  year      = {2015},
}

@Book{Sternberg2016CognitivePsychology,
  author    = {Sternberg, Robert J. and Sternberg, Karin},
  publisher = {Nelson Education},
  title     = {{Cognitive psychology}},
  year      = {2016},
}

@Article{Strobelt2018Seq2seqVisVisual,
  author    = {Strobelt, Hendrik and Gehrmann, Sebastian and Behrisch, Michael and Perer, Adam and Pfister, Hanspeter and Rush, Alexander M},
  journal   = {{IEEE} Transactions on Visualization and Computer Graphics},
  title     = {Seq2Seq-Vis: A visual debugging tool for sequence-to-sequence models},
  year      = {2018},
  number    = {1},
  pages     = {353--363},
  volume    = {25},
  publisher = {IEEE},
}

@Article{Strobelt2022GenniHumanAi,
  author    = {Hendrik Strobelt and Jambay Kinley and Robert Krueger and Johanna Beyer and Hanspeter Pfister and Alexander M. Rush},
  journal   = {{IEEE} Transactions on Visualization and Computer Graphics},
  title     = {{GenNI}: Human-{AI} Collaboration for Data-Backed Text Generation},
  year      = {2022},
  number    = {1},
  pages     = {1106--1116},
  volume    = {28},
  publisher = {Institute of Electrical and Electronics Engineers ({IEEE})},
}

@InProceedings{Tan2022EnhancingRecommendationAutomated,
  author    = {Yanchao Tan and Carl Yang and Xiangyu Wei and Chaochao Chen and Longfei Li and Xiaolin Zheng},
  booktitle = {{IEEE} 38th International Conference on Data Engineering ({ICDE})},
  title     = {Enhancing Recommendation with Automated Tag Taxonomy Construction in Hyperbolic Space},
  year      = {2022},
  publisher = {{IEEE}},
}

@Article{Teuling2010BivariateColourMaps,
  author    = {A. J. Teuling and R. Stöckli and S. I. Seneviratne},
  journal   = {International Journal of Climatology},
  title     = {Bivariate colour maps for visualizing climate data},
  year      = {2010},
  number    = {9},
  pages     = {1408--1412},
  volume    = {31},
  publisher = {Wiley},
}

@Article{Vaswani2017AttentionIsAll,
  author        = {Ashish Vaswani and Noam Shazeer and Niki Parmar and Jakob Uszkoreit and Llion Jones and Aidan N. Gomez and Lukasz Kaiser and Illia Polosukhin},
  title         = {Attention Is All You Need},
  year          = {2017},
  archiveprefix = {arXiv},
  eprint        = {1706.03762},
}

@InProceedings{Wiedemann2019DoesBertMake,
  author    = {Gregor Wiedemann and Steffen Remus and Avi Chawla and Chris Biemann},
  booktitle = {Proceedings of {KONVENS}},
  title     = {{Does {BERT} Make Any Sense? Interpretable Word Sense Disambiguation with Contextualized Embeddings}},
  year      = {2019},
  address   = {Erlangen, Germany},
}

@InProceedings{Williams2018BroadCoverageChallenge,
  author    = {Williams, Adina and Nangia, Nikita and Bowman, Samuel},
  booktitle = {Proceedings of the Conference of the North American Chapter of the Association for Computational Linguistics: Human Language Technologies},
  title     = {A Broad-Coverage Challenge Corpus for Sentence Understanding through Inference},
  year      = {2018},
  address   = {New Orleans, Louisiana},
  pages     = {1112--1122},
  publisher = {Association for Computational Linguistics},
  volume    = {1},
}

@InProceedings{Wolf2020TransformersStateArt,
  author    = {Thomas Wolf and Lysandre Debut and Victor Sanh and Julien Chaumond and Clement Delangue and Anthony Moi and Pierric Cistac and Tim Rault and Rémi Louf and Morgan Funtowicz and Joe Davison and Sam Shleifer and Patrick von Platen and Clara Ma and Yacine Jernite and Julien Plu and Canwen Xu and Teven Le Scao and Sylvain Gugger and Mariama Drame and Quentin Lhoest and Alexander M. Rush},
  booktitle = {Proceedings of the Conference on Empirical Methods in Natural Language Processing: System Demonstrations},
  title     = {Transformers: State-of-the-Art Natural Language Processing},
  year      = {2020},
  address   = {Online},
  pages     = {38--45},
  publisher = {Association for Computational Linguistics},
}

@Misc{Workshop2023Bloom176bParameter,
  author        = {Teven Le Scao and Angela Fan and Christopher Akiki and Ellie Pavlick and Suzana Ilić and Daniel Hesslow and Roman Castagné and Alexandra Sasha Luccioni and François Yvon and Matthias Gallé and Jonathan Tow and Alexander M. Rush and Stella Biderman and Albert Webson and Pawan Sasanka Ammanamanchi and Thomas Wang and Benoît Sagot and Niklas Muennighoff and Albert Villanova del Moral and Olatunji Ruwase and Rachel Bawden and Stas Bekman and Angelina McMillan-Major and Iz Beltagy and Huu Nguyen and Lucile Saulnier and Samson Tan and Pedro Ortiz Suarez and Victor Sanh and Hugo Laurençon and Yacine Jernite and Julien Launay and Margaret Mitchell and Colin Raffel and Aaron Gokaslan and Adi Simhi and Aitor Soroa and Alham Fikri Aji and Amit Alfassy and Anna Rogers and Ariel Kreisberg Nitzav and Canwen Xu and Chenghao Mou and Chris Emezue and Christopher Klamm and Colin Leong and Daniel van Strien and David Ifeoluwa Adelani and Dragomir Radev and Eduardo González Ponferrada and Efrat Levkovizh and Ethan Kim and Eyal Bar Natan and Francesco De Toni and Gérard Dupont and Germán Kruszewski and Giada Pistilli and Hady Elsahar and Hamza Benyamina and Hieu Tran and Ian Yu and Idris Abdulmumin and Isaac Johnson and Itziar Gonzalez-Dios and Javier de la Rosa and Jenny Chim and Jesse Dodge and Jian Zhu and Jonathan Chang and Jörg Frohberg and Joseph Tobing and Joydeep Bhattacharjee and Khalid Almubarak and Kimbo Chen and Kyle Lo and Leandro Von Werra and Leon Weber and Long Phan and Loubna Ben allal and Ludovic Tanguy and Manan Dey and Manuel Romero Muñoz and Maraim Masoud and María Grandury and Mario Šaško and Max Huang and Maximin Coavoux and Mayank Singh and Mike Tian-Jian Jiang and Minh Chien Vu and Mohammad A. Jauhar and Mustafa Ghaleb and Nishant Subramani and Nora Kassner and Nurulaqilla Khamis and Olivier Nguyen and Omar Espejel and Ona de Gibert and Paulo Villegas and Peter Henderson and Pierre Colombo and Priscilla Amuok and Quentin Lhoest and Rheza Harliman and Rishi Bommasani and Roberto Luis López and Rui Ribeiro and Salomey Osei and Sampo Pyysalo and Sebastian Nagel and Shamik Bose and Shamsuddeen Hassan Muhammad and Shanya Sharma and Shayne Longpre and Somaieh Nikpoor and Stanislav Silberberg and Suhas Pai and Sydney Zink and Tiago Timponi Torrent and Timo Schick and Tristan Thrush and Valentin Danchev and Vassilina Nikoulina and Veronika Laippala and Violette Lepercq and Vrinda Prabhu and Zaid Alyafeai and Zeerak Talat and Arun Raja and Benjamin Heinzerling and Chenglei Si and Davut Emre Taşar and Elizabeth Salesky and Sabrina J. Mielke and Wilson Y. Lee and Abheesht Sharma and Andrea Santilli and Antoine Chaffin and Arnaud Stiegler and Debajyoti Datta and Eliza Szczechla and Gunjan Chhablani and Han Wang and Harshit Pandey and Hendrik Strobelt and Jason Alan Fries and Jos Rozen and Leo Gao and Lintang Sutawika and M Saiful Bari and Maged S. Al-shaibani and Matteo Manica and Nihal Nayak and Ryan Teehan and Samuel Albanie and Sheng Shen and Srulik Ben-David and Stephen H. Bach and Taewoon Kim and Tali Bers and Thibault Fevry and Trishala Neeraj and Urmish Thakker and Vikas Raunak and Xiangru Tang and Zheng-Xin Yong and Zhiqing Sun and Shaked Brody and Yallow Uri and Hadar Tojarieh and Adam Roberts and Hyung Won Chung and Jaesung Tae and Jason Phang and Ofir Press and Conglong Li and Deepak Narayanan and Hatim Bourfoune and Jared Casper and Jeff Rasley and Max Ryabinin and Mayank Mishra and Minjia Zhang and Mohammad Shoeybi and Myriam Peyrounette and Nicolas Patry and Nouamane Tazi and Omar Sanseviero and Patrick von Platen and Pierre Cornette and Pierre François Lavallée and Rémi Lacroix and Samyam Rajbhandari and Sanchit Gandhi and Shaden Smith and Stéphane Requena and Suraj Patil and Tim Dettmers and Ahmed Baruwa and Amanpreet Singh and Anastasia Cheveleva and Anne-Laure Ligozat and Arjun Subramonian and Aurélie Névéol and Charles Lovering and Dan Garrette and Deepak Tunuguntla and Ehud Reiter and Ekaterina Taktasheva and Ekaterina Voloshina and Eli Bogdanov and Genta Indra Winata and Hailey Schoelkopf and Jan-Christoph Kalo and Jekaterina Novikova and Jessica Zosa Forde and Jordan Clive and Jungo Kasai and Ken Kawamura and Liam Hazan and Marine Carpuat and Miruna Clinciu and Najoung Kim and Newton Cheng and Oleg Serikov and Omer Antverg and Oskar van der Wal and Rui Zhang and Ruochen Zhang and Sebastian Gehrmann and Shachar Mirkin and Shani Pais and Tatiana Shavrina and Thomas Scialom and Tian Yun and Tomasz Limisiewicz and Verena Rieser and Vitaly Protasov and Vladislav Mikhailov and Yada Pruksachatkun and Yonatan Belinkov and Zachary Bamberger and Zdeněk Kasner and Alice Rueda and Amanda Pestana and Amir Feizpour and Ammar Khan and Amy Faranak and Ana Santos and Anthony Hevia and Antigona Unldreaj and Arash Aghagol and Arezoo Abdollahi and Aycha Tammour and Azadeh HajiHosseini and Bahareh Behroozi and Benjamin Ajibade and Bharat Saxena and Carlos Muñoz Ferrandis and Daniel McDuff and Danish Contractor and David Lansky and Davis David and Douwe Kiela and Duong A. Nguyen and Edward Tan and Emi Baylor and Ezinwanne Ozoani and Fatima Mirza and Frankline Ononiwu and Habib Rezanejad and Hessie Jones and Indrani Bhattacharya and Irene Solaiman and Irina Sedenko and Isar Nejadgholi and Jesse Passmore and Josh Seltzer and Julio Bonis Sanz and Livia Dutra and Mairon Samagaio and Maraim Elbadri and Margot Mieskes and Marissa Gerchick and Martha Akinlolu and Michael McKenna and Mike Qiu and Muhammed Ghauri and Mykola Burynok and Nafis Abrar and Nazneen Rajani and Nour Elkott and Nour Fahmy and Olanrewaju Samuel and Ran An and Rasmus Kromann and Ryan Hao and Samira Alizadeh and Sarmad Shubber and Silas Wang and Sourav Roy and Sylvain Viguier and Thanh Le and Tobi Oyebade and Trieu Le and Yoyo Yang and Zach Nguyen and Abhinav Ramesh Kashyap and Alfredo Palasciano and Alison Callahan and Anima Shukla and Antonio Miranda-Escalada and Ayush Singh and Benjamin Beilharz and Bo Wang and Caio Brito and Chenxi Zhou and Chirag Jain and Chuxin Xu and Clémentine Fourrier and Daniel León Periñán and Daniel Molano and Dian Yu and Enrique Manjavacas and Fabio Barth and Florian Fuhrimann and Gabriel Altay and Giyaseddin Bayrak and Gully Burns and Helena U. Vrabec and Imane Bello and Ishani Dash and Jihyun Kang and John Giorgi and Jonas Golde and Jose David Posada and Karthik Rangasai Sivaraman and Lokesh Bulchandani and Lu Liu and Luisa Shinzato and Madeleine Hahn de Bykhovetz and Maiko Takeuchi and Marc Pàmies and Maria A Castillo and Marianna Nezhurina and Mario Sänger and Matthias Samwald and Michael Cullan and Michael Weinberg and Michiel De Wolf and Mina Mihaljcic and Minna Liu and Moritz Freidank and Myungsun Kang and Natasha Seelam and Nathan Dahlberg and Nicholas Michio Broad and Nikolaus Muellner and Pascale Fung and Patrick Haller and Ramya Chandrasekhar and Renata Eisenberg and Robert Martin and Rodrigo Canalli and Rosaline Su and Ruisi Su and Samuel Cahyawijaya and Samuele Garda and Shlok S Deshmukh and Shubhanshu Mishra and Sid Kiblawi and Simon Ott and Sinee Sang-aroonsiri and Srishti Kumar and Stefan Schweter and Sushil Bharati and Tanmay Laud and Théo Gigant and Tomoya Kainuma and Wojciech Kusa and Yanis Labrak and Yash Shailesh Bajaj and Yash Venkatraman and Yifan Xu and Yingxin Xu and Yu Xu and Zhe Tan and Zhongli Xie and Zifan Ye and Mathilde Bras and Younes Belkada and Thomas Wolf},
  title         = {BLOOM: A 176B-Parameter Open-Access Multilingual Language Model},
  year          = {2023},
  archiveprefix = {arXiv},
  eprint        = {2211.05100},
}

@InProceedings{Xiang2021OntoeaOntologyGuided,
  author    = {Yuejia Xiang and Ziheng Zhang and Jiaoyan Chen and Xi Chen and Zhenxi Lin and Yefeng Zheng},
  booktitle = {Findings of the Association for Computational Linguistics: {ACL}-{IJCNLP}},
  title     = {{OntoEA}: Ontology-guided Entity Alignment via Joint Knowledge Graph Embedding},
  year      = {2021},
  publisher = {Association for Computational Linguistics},
}

@InProceedings{Xu2022TaxopromptPromptBased,
  author    = {Hongyuan Xu and Yunong Chen and Zichen Liu and Yanlong Wen and Xiaojie Yuan},
  booktitle = {Proceedings of the 31st International Joint Conference on Artificial Intelligence},
  title     = {{TaxoPrompt}: A Prompt-based Generation Method with Taxonomic Context for Self-Supervised Taxonomy Expansion},
  year      = {2022},
  publisher = {International Joint Conference on Artificial Intelligence Organization},
}

@Article{Yu2022SurveyKnowledgeEnhanced,
  author    = {Wenhao Yu and Chenguang Zhu and Zaitang Li and Zhiting Hu and Qingyun Wang and Heng Ji and Meng Jiang},
  journal   = {{ACM} Computing Surveys},
  title     = {A Survey of Knowledge-enhanced Text Generation},
  year      = {2022},
  number    = {11s},
  pages     = {1--38},
  volume    = {54},
  publisher = {Association for Computing Machinery ({ACM})},
}

@InProceedings{Yuan2022WordcraftStoryWriting,
  author    = {Ann Yuan and Andy Coenen and Emily Reif and Daphne Ippolito},
  booktitle = {27th International Conference on Intelligent User Interfaces},
  title     = {Wordcraft: Story Writing With Large Language Models},
  year      = {2022},
  publisher = {{ACM}},
}

@InProceedings{Zhang2018Taxogen,
  author    = {Chao Zhang and Fangbo Tao and Xiusi Chen and Jiaming Shen and Meng Jiang and Brian Sadler and Michelle Vanni and Jiawei Han},
  booktitle = {Proceedings of the 24th {ACM} {SIGKDD} International Conference on Knowledge Discovery and Data Mining},
  title     = {{TaxoGen}},
  year      = {2018},
  publisher = {{ACM}},
}

@Article{Zhang2022SurveyControllableText,
  author        = {Hanqing Zhang and Haolin Song and Shaoyu Li and Ming Zhou and Dawei Song},
  title         = {A Survey of Controllable Text Generation using Transformer-based Pre-trained Language Models},
  year          = {2022},
  archiveprefix = {arXiv},
  eprint        = {2201.05337},
}

@Article{Zhao2024ExplainabilityLargeLanguage,
  author    = {Zhao, Haiyan and Chen, Hanjie and Yang, Fan and Liu, Ninghao and Deng, Huiqi and Cai, Hengyi and Wang, Shuaiqiang and Yin, Dawei and Du, Mengnan},
  journal   = {ACM Transactions on Intelligent Systems and Technology},
  title     = {Explainability for Large Language Models: A Survey},
  year      = {2024},
  issn      = {2157-6912},
  publisher = {Association for Computing Machinery (ACM)},
}
